\documentclass{article}
\usepackage{arxiv}

\usepackage[utf8]{inputenc} 
\usepackage[T1]{fontenc}    
\usepackage{hyperref}       
\usepackage{url}            
\usepackage{booktabs}       
\usepackage{amsfonts}       
\usepackage{nicefrac}       
\usepackage{microtype}      
\usepackage{lipsum}

\usepackage{array}
\usepackage{amsmath}
\usepackage[pdftex]{graphicx}
\graphicspath{{fig/pdf/}}
\DeclareGraphicsExtensions{.pdf}
\usepackage{import}
\usepackage{adjustbox}
\usepackage[caption=false,font=normalsize,labelfont=sf,textfont=sf]{subfig}

\usepackage{multirow}

\usepackage{amssymb}
\usepackage{mathtools}
\newcommand{\rev}[1]{{#1}}

\title{Eye Know You: Metric Learning for End-to-end Biometric Authentication Using Eye Movements from a Longitudinal Dataset}

\author{Dillon~Lohr\\
Department of Computer Science\\
Texas State University\\
San Marcos, TX 78666 USA\\
\texttt{djl70@txstate.edu}\\
\And
Henry~Griffith\\
Department of Computer Science\\
Texas State University\\
San Marcos, TX 78666 USA\\
\texttt{hkgriffith1@gmail.com}\\
\And
Oleg~V~Komogortsev\\
Department of Computer Science\\
Texas State University\\
San Marcos, TX 78666 USA\\
\texttt{ok11@txstate.edu}\\
}

\begin{document}
\maketitle

\begin{abstract}
The permanence of eye movements as a biometric modality remains largely unexplored in the literature.
The present study addresses this limitation by evaluating a novel exponentially-dilated convolutional neural network for eye movement authentication using a recently proposed longitudinal dataset known as GazeBase.
The network is trained using multi-similarity loss, which directly enables the enrollment and authentication of out-of-sample users.
In addition, this study includes an exhaustive analysis of the effects of evaluating on various tasks and downsampling from 1000~Hz to several lower sampling rates.
Our results reveal that reasonable authentication accuracy may be achieved even during both a low-cognitive-load task and at low sampling rates.
Moreover, we find that eye movements are quite resilient against template aging after as long as 3~years.
\end{abstract}

\section{Introduction}

Eye movement biometrics have received considerable attention in the literature over the past two decades~\cite{Katsini2020}.
This focus is motivated by the specificity and permanence of human eye movements~\cite{Bargary2017}.
Eye movement biometric systems offer notable advantages over alternative modalities, including the ability to support liveness detection~\cite{Komogortsev2015, Makowski2020} and spoof-resistant continuous authentication~\cite{Eberz2015}.
Eye movements are also well suited for integration within multimodal biometric systems~\cite{Kasprowski2018}.

Despite the considerable literature within this domain, several improvements are necessary to advance the large-scale commercial viability of this technology.
For example, most existing literature formulates eye movement biometrics as a closed-set classification problem~\cite{George2016, Kasprowski2004, Li2018, Jia2018}.
This approach is problematic for real-world scenarios in which new users must be continuously enrolled and authenticated. 

Moreover, the existing knowledge base is further limited by the validation of the proposed models on a variety of diverse datasets, many of which are characterized by short-term test-retest intervals~\cite{Kasprowski2004, Li2018, Abdelwahab2019, Lohr2020}.
This lack of a suitable gold-standard validation set with sufficient temporal duration limits both comparability between results and the assessment of template aging effects.
Finally, although eye tracking sensors within emerging commercial devices often are characterized by limited temporal precision and deployment in low resource environments, few studies have explored performance variability versus signal sampling rate, nor the capacity to reduce the number of model parameters to support deployment in embedded environments. 

The research described herein attempts to address many of the aforementioned limitations.
We train an exponentially-dilated convolutional neural network~(CNN) that learns meaningful embeddings via multi-similarity~(MS) loss~\cite{Wang2019}.
Inputs consist of fixed-length subsequences of eye movements during various tasks, including reading, tracking jumping dots, watching videos, and playing an interactive game.
Similarity scores are measured as the mean cosine similarity between temporally-aligned subsequence embeddings.
The proposed technique is verified on several tasks from the GazeBase dataset~\cite{Griffith2020}, which consists of 322~participants recorded up to 18~times each over a 37-month period.
We also compare against a statistical baseline and the current state-of-the-art, DeepEyedentificationLive~(DEL)~\cite{Makowski2020}.

The main contributions of this study are:
\begin{itemize}
    \item
    \rev{The development of an exponentially-dilated convolutional neural network model offering state-of-the-art performance with 440 times fewer learnable parameters.}
    \item
    The initial demonstration of multi-similarity loss in a metric learning framework for eye movement biometrics.
    \item
    The most thorough assessment of eye movement permanence to date, with reasonable authentication performance demonstrated for a 37-month test-retest interval.
    \item
    The most thorough assessment of task dependence to date, with comparable authentication performance achieved for a low-cognitive-load task (i.e., jumping dot stimulus) versus traditionally recommended high-cognitive-load tasks (e.g., reading~\cite{Holland2011} or visual search~\cite{Li2018}).
\end{itemize}

\section{Prior Work}
\begin{table*}
    \centering
    \caption{A summary of the methodological aspects of selected works.
    N is the number of subjects employed when measuring performance; the full size of the dataset may have been larger.
    ST means short-term and LT means long-term. \\
    *: dataset was previously public but is unavailable at the time of writing. \\
    **: a modified version or subset of the dataset is publicly available.}
    \label{tab:prior_work}
    \begin{adjustbox}{width=\textwidth}
    \begin{tabular}{lllp{4cm}p{4cm}p{2cm}p{3.5cm}l}
        \toprule
        Study & Year & Open-set? & Tasks & Sampling rates (Hz) & N subjects & Test-retest interval & Public dataset? \\
        \midrule
        \cite{Kasprowski2004} & 2004 & N & Jumping dot & 250 & 9 & same day & N \\
        \midrule
        \cite{George2016} & 2015 & N & Reading; jumping dot & 250 & \parbox[t]{2cm}{76--77 (ST);\\ 18--19 (LT)} & \parbox[t]{3.5cm}{30~min. (ST);\\ 1~year (LT)} & N* \\
        \midrule
        \cite{Friedman2017} & 2017 & Y & Reading & 1000 & \parbox[t]{2cm}{149 (ST);\\ 34 (LT)} & \parbox[t]{3.5cm}{med. 19~min. (ST);\\med. 11.1~months (LT)} & Y**~\cite{Griffith2020} \\
        \midrule
        \cite{Li2018} & 2018 & N & Visual search & 300, 150, 75, 30 & 58 & \parbox[t]{3.5cm}{same day (ST);\\ avg. 18~days (LT)} & N \\
        \midrule
        \cite{Jia2018} & 2018 & N & Image viewing & 500 & 32 & $\leq$~30~min. & N \\
        \midrule
        \cite{Abdelwahab2019} & 2019 & Y & Video viewing & 30 & 105 & same day & Y~\cite{Mital2011} \\
        \midrule
        \cite{Jager2020} & 2020 & Y & \parbox[t]{4cm}{Reading (ST);\\ jumping dot (LT)} & 1000 & \parbox[t]{2cm}{25 (ST);\\ 10 (LT)} & \parbox[t]{3.5cm}{same day (ST);\\ 2--8 weeks (LT)} & Y**~\cite{Jager2020} \\
        \midrule
        \cite{Makowski2020} & 2020 & Y & Jumping dot & 1000 & 25 & $\geq$~1--4~weeks & Y~\cite{Makowski2020} \\
        \midrule
        \cite{Prasse2020} & 2020 & Y & Jumping dot & \parbox[t]{4cm}{1000, 500, 250, 125, 62, 31} & 25 & $\geq$~1--4~weeks & Y~\cite{Makowski2020} \\
        \midrule
        \cite{Lohr2020} & 2020 & Y & Reading & 1000 & 67--68 & avg. 20~min. & Y**~\cite{Griffith2020} \\
        \midrule
        Present & 2021 & Y & \parbox[t]{4cm}{Reading; jumping dot;\\ static dot; video viewing;\\ interactive game} & \parbox[t]{4cm}{1000 (all tasks);\\ 500, 250, 125, 50, 31.25\\ (reading only)} & 14--59 & 20~min. to 37~months & Y~\cite{Griffith2020} \\
        \bottomrule
    \end{tabular}
    \end{adjustbox}
\end{table*}

Since the introduction of eye movements as a biometric in 2004~\cite{Kasprowski2004}, significant research has focused on improving their viability.
A collective review of related work published prior to 2015 may be found in~\cite{Galdi2016}.
Moreover, comparative results for studies analyzing common datasets are provided in~\cite{Rigas2017}, which summarizes the results of the most recent BioEye competition.
As noted within these reviews, the majority of prior work uses a common processing pipeline, with the recordings initially partitioned into specific eye movement events using a classification algorithm, followed by the formation of the biometric template as a vector of discrete features from each event.
One problem with such approaches is that event classification is a difficult problem~\cite{Andersson2017}, so it adds another layer of complexity that influences biometric performance.
Only recently have studies begun utilizing end-to-end deep learning workflows ~\cite{Jia2018, Abdelwahab2019}.

The winners of the BioEye 2015 competition, George \& Routray~\cite{George2016}, used a radial basis function~(RBF) network for computing similarities between probe and gallery vectors.
Features describing the position, velocity, and acceleration for fixations and saccades were extracted from the segmented signal.
The algorithm was validated using a dataset of 153~individuals recorded twice during both a reading task (TEX) and a random saccades task (RAN) with 30~minutes between recording sessions and recorded again after one year.
They achieved an equal error rate~(EER) of~2.59\% for RAN and 3.78\% for TEX when the recording sessions were separated by 30~minutes.
When the recording sessions were separated by one year, they achieved 10.96\% EER for RAN and 9.36\% for TEX.
As the proposed method requires retraining the network upon the enrollment of each new user, it is not feasible for large-scale practical deployment.

In addition to eye movement-specific features, other representations of eye movement recordings have also been explored in the literature.
For example, Li et al.~\cite{Li2018} used a multi-channel Gabor wavelet transform~(GWT) to extract texture features from eye movement trajectories during a visual search task.
Support vector machine~(SVM) classifiers were used for biometric identification and verification.
Results were verified using a dataset consisting of 58~subjects recorded across several trials, with a minimum EER of~0.89\% reported.
Texture-based eye movement features were recently reinvestigated in~\cite{Griffith2020b}, where downsampling of the filtered images was proposed for the feature extraction step in order to preserve spatial structure.
In addition to the aforementioned restriction regarding new user enrollment, both of these studies utilized recordings with only a small temporal separation.

Jia et al.~\cite{Jia2018} introduced deep learning techniques for eye movement biometrics.
A recurrent neural network~(RNN) was built using long short-term memory~(LSTM) cells.
The output layer used softmax to produce class probabilities.
Their approach was validated using a dataset of 32~subjects recorded across several trials of a high-cognitive-load task, with a minimum EER of~0.85\% reported.
This study did not explore its method's long-term efficacy, as recordings for each subject were collected during a single, 30-minute period.

Friedman et al.~\cite{Friedman2017} employed a statistical approach for eye movement biometrics.
A novel event classification algorithm, the modified Nystr\"{o}m and Holmqvist~(MNH) algorithm~\cite{Friedman2018}, was used to classify several types of events.
A set of over 1,000~features~\cite{Rigas2018} was extracted from each recording.
This approach was validated using a subset of the dataset considered herein, consisting of 298~subjects recorded twice each during a reading task. 
Using data separated by approximately 20~minutes, a best-case EER of~2.01\% was reported.
With data separated by approximately 11~months from a set of 68~subjects, EER increased to~10.16\%.

J\"{a}ger et al.~\cite{Jager2020} utilized involuntary micro eye movements for biometric authentication and identification.
Raw eye movement signals were initially transformed to isolate desired micro eye movements according to their characteristic velocities, with the resulting scaled values fed into a CNN with two separate subnets.
The approach was validated using two datasets (75~subjects during a reading task recorded at 1000~Hz~\cite{Makowski2018}, and a newly recorded dataset consisting of 10~users).
This approach was later extended into DeepEyedentificationLive~(DEL)~\cite{Makowski2020} to include liveness detection and was evaluated on a different dataset of 150~subjects, the JuDo1000 dataset~\cite{Makowski2020}, which is publicly available.
However, the EERs presented in the later study were based on only 25~identities, and the recordings were collected with a relatively short temporal separation.
DEL was also evaluated on temporally- and spatially-degraded signals by Prasse et al.~\cite{Prasse2020}.

Abdelwahab \& Landwehr~\cite{Abdelwahab2019} introduced metric learning to the eye movement biometrics literature using deep distributional embeddings.
Namely, sequences of six-dimensional vectors (binocular gaze and pupil data) at 30~Hz were fed to a deep neural network which produced distributional embeddings using a Wasserstein distance metric.
The approach was validated on the publicly-available Dynamic Images and Eye Movements~(DIEM) dataset~\cite{Mital2011}, which contains eye movement data of 210~subjects viewing various video clips (sports, movie trailers, etc.).
The recordings in the DIEM dataset were collected with only a small temporal separation.

Lohr et al.~\cite{Lohr2020} also explored the use of metric learning for eye movement biometrics.
Eye movement recordings were segmented into fixations, saccades, and PSOs using the MNH algorithm~\cite{Friedman2018}, and discrete feature vectors were extracted from each event.
Three separate multilayer perceptrons~(MLPs), one for each of the 3~event types, were trained on these feature vectors with triplet loss~\cite{Schroff2015} to create meaningful embeddings.
Distances were computed for each event type separately and then fused with a weighted sum.
The approach was validated using a dataset of 269~subjects recorded twice each during a reading task.
An average EER of~6.29\% was reported for recordings separated by approximately 20~minutes.
Like most prior studies, the permanence of eye movements was not explored.

The technique described herein expands upon the work of Lohr et al.~\cite{Lohr2020} by feeding recordings directly into the model (removing the additional complexity of event classification).
Additionally, the more sophisticated MS loss~\cite{Wang2019} is used, a single exponentially-dilated CNN is trained rather than multiple event-specific MLPs, and performance is evaluated on a longitudinal dataset collected over a 37-month period.
The present study also explores the authentication performance of additional tasks other than reading and of downsampled eye movement signals.

\delimitershortfall=-1pt

\section{Methodology}

\subsection{Dataset}
We used the GazeBase~\cite{Griffith2020} dataset available on Figshare~\cite{Griffith2020a}.
This dataset consists of 322~college-aged subjects, each recorded monocularly (left eye only) at 1000~Hz with an EyeLink~1000 eye tracker.
Nine rounds of recordings (R1--9) were captured over a period of 37~months, thereby enabling the analysis of template aging.
Each subsequent round comprises a subset of subjects from the preceding round (with one exception, subject~76, who was absent from R3 but returned for R4--5), with only 14~of the initial 322~subjects present across all 9~rounds.
Each round consists of 2~recording sessions separated by approximately 30~minutes, totaling 18~recording sessions.
Recordings contain the horizontal and vertical components of the left eye's gaze position in terms of degrees of the visual angle.
In each recording session, every subject performed a series of~7 eye movement tasks: a horizontal saccades task~(HSS), a video-viewing task~(VD1), a fixation task~(FXS), a random saccades task~(RAN), a reading task~(TEX), a ball-popping task~(BLG), and another video-viewing task~(VD2).
More details for each task can be found in~\cite{Griffith2020}.
Since VD2 was similar to VD1, we only used VD1 in our experiments.

\subsection{Training and testing splits}
The subjects in the dataset were split in the following manner.
First, we created a held-out test set using all recordings from the 59~subjects that were present in R6.
The test set contained nearly 50\% of all recordings in GazeBase.
The test set was only used at the very end of our experiments to get a final, unbiased measure of our models' performance.

Next, we split the remaining subjects into 4~folds (which we will label F1--4) for cross-validation.
We had three goals when balancing the folds, keeping in mind that some subjects have more recordings than others: (1)~each fold should have a similar number of subjects, (2)~each fold should have a similar number of recordings, and (3)~the method to create the folds should be deterministic to facilitate reproducibility.

We accomplished these goals by using two priority queues (heaps)---one for the folds and the other for the subjects---and iteratively assigning subjects to folds.
Each fold was weighted first by the number of subjects assigned to it and second by the total number of recordings present for those subjects, and the fold with the lowest weight was given the highest priority.
Each subject was weighted by the total number of recordings present for that subject, and the subject with the highest weight was given the highest priority.
In case of ties, an arbitrary-but-deterministic element was given higher priority.
At each iteration, we extracted the highest priority element from both heaps and assigned the chosen subject to the chosen fold.
The chosen fold was then placed back onto the heap with its updated priority.
This process was repeated until the subject heap was empty.
In the end, the largest fold had at most 1~more subject than the smallest fold, and the number of recordings present in each fold was as balanced as possible.

We ran three sets of experiments: (1)~compare our metric learning model against three baseline models, using data from TEX and tuning hyperparameters based on the average performance across the 4~folds; (2)~use data from the other individual tasks to assess our model's performance on types of eye movements other than reading; and (3)~downsample the TEX data to assess our model's performance on signals with lower sampling rates.

\subsection{Signal pre-processing}
We start with a sequence of $T$ tuples $(t^{(i)},x^{(i)},y^{(i)}),\, i = 1, \dotsc, T$, where $t^{(i)}$ is the time stamp (s) and $x^{(i)},y^{(i)}$ are the horizontal and vertical components of the monocular (left eye) gaze position ($^{\circ}$).
Next, we compute per-channel velocity ($^{\circ}$/s) using the one-sample backward difference method:
\begin{align}
    \delta_x^{(i)} &= \frac{x^{(i)} - x^{(i - 1)}}{t^{(i)} - t^{(i - 1)}},\, i = 2, \dotsc, n\\
    \delta_y^{(i)} &= \frac{y^{(i)} - y^{(i - 1)}}{t^{(i)} - t^{(i - 1)}},\, i = 2, \dotsc, n.
\end{align}
Then, we replace any NaN velocities with 0 and clip both $\delta_x^{(i)}$ and $\delta_y^{(i)}$ within the range $[-1000,1000]$ to minimize the influence of outliers.

During some preliminary experiments, we found that the slow and fast velocity transformations used in the DeepEyedentification line of work~\cite{Jager2020,Prasse2020,Makowski2020} indeed led to better results on the validation set compared to z-score transforming raw velocity values.
Therefore, we extracted the following 4~values for each $(\delta_x^{(i)},\delta_y^{(i)})$ tuple:
\begin{align}
    x_{\text{slow}}^{(i)} &= \tanh(c \delta_x^{(i)})
    \label{eq:xslow} \\
    y_{\text{slow}}^{(i)} &= \tanh(c \delta_y^{(i)})
    \label{eq:yslow} \\
    x_{\text{fast}}^{(i)} &= 
    \begin{dcases}
        z_x(\delta_x^{(i)}), & \text{if } \sqrt{{\delta_x^{(i)}}^2 + {\delta_y^{(i)}}^2} \ge v \\
        z_x(0), & \text{otherwise}
    \end{dcases}
    \label{eq:xfast} \\
    y_{\text{fast}}^{(i)} &= 
    \begin{dcases}
        z_y(\delta_y^{(i)}), & \text{if } \sqrt{{\delta_x^{(i)}}^2 + {\delta_y^{(i)}}^2} \ge v \\
        z_y(0), & \text{otherwise}
    \end{dcases},
    \label{eq:yfast}
\end{align}
where $c = 0.02$ and $v = 40$ are fixed hyperparameters taken from \cite{Makowski2020}, and $z_x$ and $z_y$ are separate z-score transformations for each velocity channel.
The mean and standard deviation for $z_x$ and $z_y$ were computed across all velocities in the train set ($\delta_x^{(i)}$ and $\delta_y^{(i)}$, respectively).

\subsection{Sampling rate degradation}
The GazeBase dataset contains recordings of very high signal quality~\cite{Lohr2019} recorded with an EyeLink~1000 eye tracker.
Other eye trackers, such as Magic Leap One~\cite{ML1} or Vive Pro Eye~\cite{ViveProEye}, have lower signal quality (e.g., 60~Hz sampling rate for the Magic Leap One or 120~Hz for the Vive Pro Eye).
It is expected that in the future, eye tracking would become ubiquitous in virtual- and augmented-reality head-mounted displays due to the many benefits it could bring, including foveated rendering, continuous authentication, and increased immersion in video games.
But since eye tracking signal quality varies across devices, in this study, it was important to consider how signal quality (specifically, sampling rate) impacts authentication performance.
To this end, we downsampled the eye movement signals using SciPy's \texttt{decimate} function~\cite{SciPy2020}.
We targeted degraded sampling rates of 500, 250, 125, 50, and 31.25~Hz.
We chose 31.25~Hz instead of 30~Hz to simplify the downsampling process.

\subsection{Network architecture}
As input, we feed in a number of time steps equivalent to 1.024~s (rounded down, if not a whole number) and, if necessary, zero-pad the end of the sequence to length~1024.
For each time step, we use the 4~channels defined in Equations~\ref{eq:xslow}--\ref{eq:yfast}.
Our network performs a mapping $f: \mathbb{R}^{4\times1024} \to \mathbb{R}^{128}$.

\begin{figure}
    \centering
    \includegraphics[width=\linewidth]{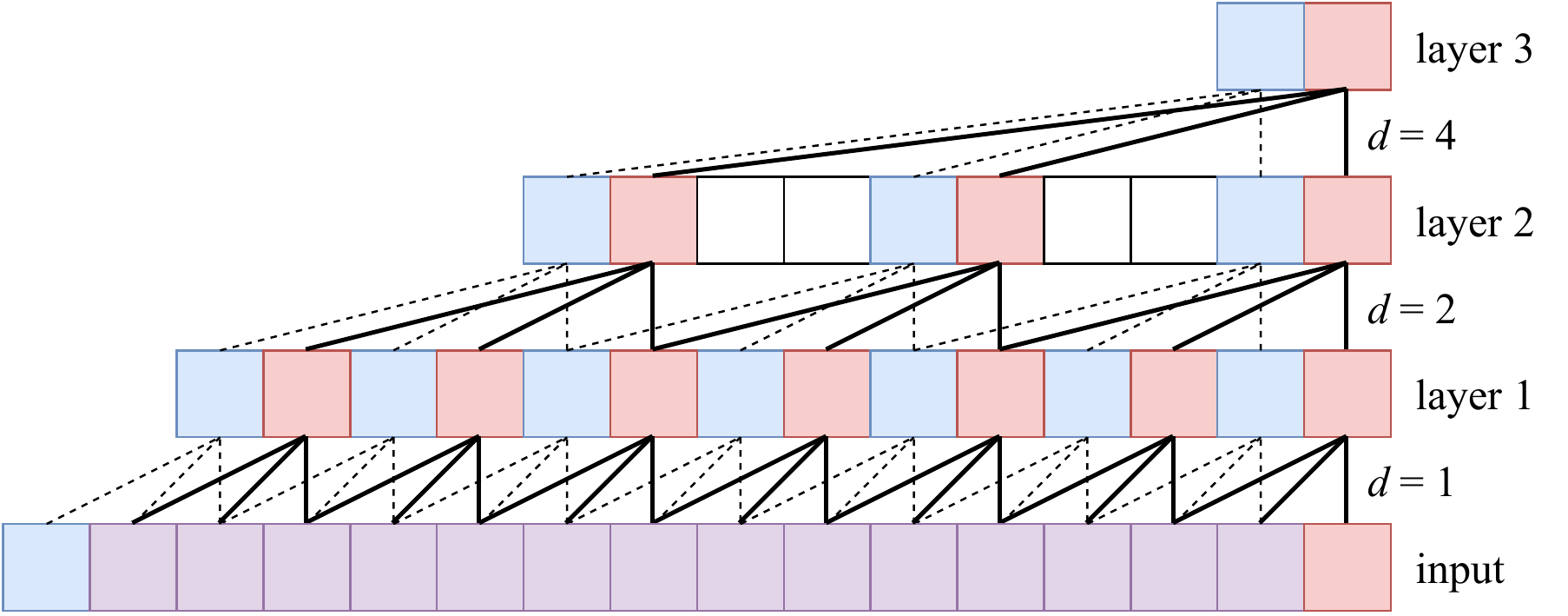}
    \caption{%
    Visualization of exponentially-dilated convolutions with kernel size~3, stride~1, and no padding.
    The convolutions in the $\ell$-th layer use a dilation of $d=2^{\ell-1}$.
    With this configuration, if the input has length $2^q$ and there are $q - 1$ layers, then the final layer has a receptive field of $2^q-1$ values from the input (shown in red and blue, with overlap in purple).%
    }
    \label{fig:exp_dilation}
\end{figure}

\begin{figure*}
    \centering
    \includegraphics[width=\linewidth]{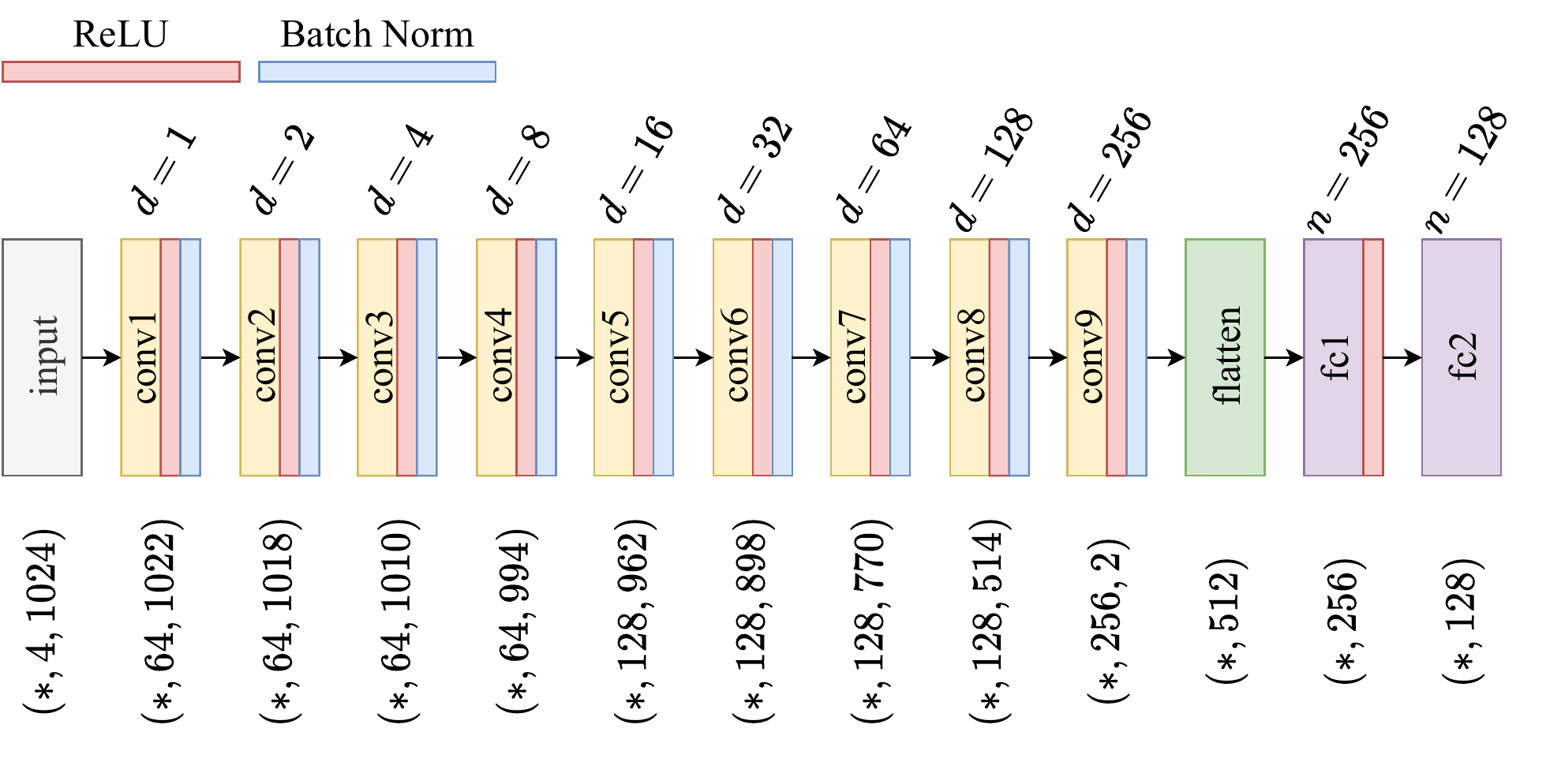}
    \caption{%
    Network architecture.
    Each convolution layer uses kernel size~3, stride~1, and no padding, and is followed by ReLU and batch normalization.
    The first fully-connected layer is followed by ReLU.
    The output of the final fully-connected layer acts as the embedding of the input.
    The numbers at the bottom reflect the shape of the data leaving each layer, ordered as minibatch size, channels, and time steps.
    The $d$ above each convolution block is the dilation used in that layer, and the $n$ above each fully-connected block is the number of output nodes in that layer.
    This network has a total of 475,264~learnable parameters.%
    }
    \label{fig:network_diagram}
\end{figure*}

The network consists of a series of exponentially-dilated convolutional layers followed by a series of fully-connected layers.
Exponentially-dilated convolutions were first proposed for semantic segmentation of images in~\cite{Yu2016} and have since seen success in other domains like audio synthesis~\cite{Oord2016} and time series classification~\cite{Bai2018,Franceschi2020}.
By exponentially increasing the dilation for subsequent convolutional layers, we achieve an exponential increase in the receptive field with only a linear increase in the number of learnable parameters.
See Figure~\ref{fig:exp_dilation} for a visualization of exponentially-dilated convolutions and Figure~\ref{fig:network_diagram} for a diagram of our network architecture.

\subsection{Multi-similarity loss}
MS loss~\cite{Wang2019}, like many other metric learning loss functions, is embedding-based (in contrast to classification-based losses like cross-entropy)~\cite{Musgrave2020} and pair-based.
Minibatches are constructed from multiple samples each from a subset of subjects so that both inter- and intra-class variations can be observed.
Pairs of samples are constructed within each minibatch.
A pair is \textit{positive} if samples in the pair are from the same class or \textit{negative} if they are from different classes.
The goal is to bring positive pairs closer together in the embedding space and to push negative pairs farther apart.
In other words, we want to construct a well-clustered embedding space.
One challenge with pair-based losses is selecting the most informative pairs to accelerate learning.
Using pairs that are too easy does not help the model learn, and using pairs that are too hard may lead to instability during training~\cite{Hermans2017}.

MS loss takes into account three different types of similarities: self-similarity, positive relative similarity, and negative relative similarity.
This is a more sophisticated approach than most other losses that focus on either self-similarity (e.g., contrastive loss~\cite{Hadsell2006}) or relative similarity (e.g., triplet loss~\cite{Schroff2015}) but not both.
The most informative pairs are selected with an online pair mining technique and assigned similarity-based weights that decay exponentially as the pairs become less informative.
A larger weight is given to positive pairs with low similarity and to negative pairs with high similarity.
Together, these aspects help MS loss form a well-clustered embedding space and overcome the challenge of selecting informative pairs.
MS loss is formulated as
\begin{align}
\begin{split}
    L = \frac{1}{m} \sum_{i=1}^{m} &\left( \frac{1}{\alpha} \log \left( 1 + \sum_{k \in P_i}{{\rm e}^{-\alpha(S_{ik} - \lambda)}} \right) \right. \\
    &+ \left. \frac{1}{\beta} \log \left( 1 + \sum_{k \in N_i}{{\rm e}^{\beta(S_{ik} - \lambda)}} \right) \!\right),
\end{split}
\end{align}
where $\alpha,\beta,\lambda$ are hyperparameters, $m$ is the size of the minibatch, $P_i$ and $N_i$ are the sets of indices of the mined positive and negative pairs for each anchor sample $\mathbf{x}_i$, and $S_{ik}$ is the cosine similarity between the pair of samples $\{\mathbf{x}_i, \mathbf{x}_k\}$.
For more technical details and descriptive figures, please refer to the MS loss paper~\cite{Wang2019}.

Although the latest metric learning loss functions (such as MS loss) may not improve upon earlier loss functions as much as the literature would suggest~\cite{Musgrave2020}, many (including MS loss) do appear to lead to marginal improvements over cross-entropy or triplet loss after controlling for several factors including network architecture, batch size, and optimizer.

\subsection{Measuring similarity between two recordings}
\label{sec:similarity}

The similarity between two recordings is measured as the mean cosine similarity across the first $n$ temporally-aligned subsequence embeddings.
That is, given two recordings $A,B$ where $\mathbf{a}^{(i)}$ is the embedding of the $i$-th non-overlapping subsequence from $A$ (and $\mathbf{b}^{(i)}$ from $B$), we compute the similarity between $A$ and $B$, denoted $S_{A,B}$, as
\begin{equation}
    \label{eq:sim}
    S_{A,B} = \frac{1}{n}\sum_{i=1}^{n}{\frac{\mathbf{a}^{(i)} \cdot \mathbf{b}^{(i)}}{\lvert\lvert\mathbf{a}^{(i)}\rvert\rvert \, \lvert\lvert\mathbf{b}^{(i)}\rvert\rvert}}.
\end{equation}

\subsection{Training}
\label{sec:training}
For a given task and sampling rate, we trained 4~different models, each one using a different held-out fold as the validation set and the remaining 3~folds as the training set.

We used the AdamW~\cite{Loshchilov2019} optimizer with learning rate and weight decay determined via hyperparameter search (see Section~\ref{sec:opt}).
All other optimizer hyperparameters were left at their default values.
We used MS loss~\cite{Wang2019} with an online miner as implemented in the PyTorch Metric Learning~(PML) library~\cite{Musgrave2020a}.
The hyperparameters for MS loss were also determined via hyperparameter search.

When constructing a minibatch of size $m$, we observed that simply selecting $k$~samples each from $\frac{m}{k}$~classes/subjects tended to oversample from R1.
To sample from all recording rounds equally and to guarantee that each minibatch contains longitudinal similarities, we pick one subject present in all of R1--5 (i.e., all rounds present in the training/validation sets), along with another random subject from each of R1--5.
Then, for each of R1--5, we randomly sample $k$~subsequences each from the two subjects chosen for that round, with half of those samples taken from recording session~1 and the other half from recording session~2.
Since we have 2~subjects each from 5~rounds, our minibatch size is $m = 10k$.
We set $k=8$ for a minibatch size of~80.

Rather than sampling from a fixed set of subsequences (e.g., constructed with a rolling window approach), the subsequences used during training were chosen from arbitrary positions in each recording.
That is, when sampling a subsequence of length $w$ from a recording of length $T$, we start from a random time step $i \in [1,\, T-w+1]$ and use the next $w$ contiguous time steps $\{i,\, i + 1,\, \dotsc,\, i + w - 1\}$.
In doing so, we force the network to learn similarities (and differences) between arbitrary subsequences of eye movements during the task and across recording sessions, hopefully improving its ability to generalize to new subjects.
In our experiments, subsequence length $w$ was a number of time steps equivalent to 1.024~s (rounded down, if not a whole number).

After every 100~training iterations (i.e., 100~minibatches), the model's performance was evaluated in the following manner using data from the validation set.
We first computed the embeddings of the first 10~non-overlapping subsequences~(10.240~s) of each recording in session~1 of R1.
These embeddings acted as the enrollment set.
We did the same for each recording in session~2 of R1 to construct the authentication set.
We then computed $S_{A,B}$ for all $A$ in the enrollment set and all $B$ in the authentication set using Equation~\ref{eq:sim} with $n=10$.
A receiver operating characteristic~(ROC) curve was built from these similarities and was used to find the equal error rate~(EER)---the point where the false acceptance rate~(FAR) was equal to the false rejection rate~(FRR).
We repeated this process with the same enrollment set but different authentication sets built from session~2 of R2--5, resulting in a total of 5~measures of EER.
The mean of these 5~EERs was used as the combined measure of the model's performance.

The model trained for a maximum of 100,000~iterations, but we also employed early stopping to help reduce training time.
Early stopping with a patience of 200~performance evaluations (20,000~iterations)---seeking to minimize the aforementioned performance measure---was used to determine if training should stop early.
Whether training lasted the full 100,000~iterations or stopped early, we kept the version of the model with the best performance evaluation and discarded any other model checkpoints.

\subsection{Hyperparameter tuning}
\label{sec:opt}
We had 6~hyperparameters to tune: 2~for AdamW (learning rate and weight decay), 3~for MS loss ($\alpha$, $\beta$, and $\lambda$), and 1~for the online mining used within MS loss ($\varepsilon$).
Bayesian optimization~\cite{Mockus1978} was employed to optimize these hyperparameters.

A total of 31~search iterations were performed.
The first iteration used a fixed hyperparameter configuration that we empirically found to work well on the training and validation sets during preliminary experiments.
The next 5~iterations (2--6) randomly explored the search space to help prevent the optimizer from getting stuck in local optima (a fixed random seed was used so that the same 5~random points were probed for all models).
The final 25~iterations (7--31) intelligently balanced exploring and exploiting the search space using the upper confidence bound~(UCB) utility/acquisition function.

At each search iteration, a total of 4~models were trained with the selected hyperparameters using the training procedure described in Section~\ref{sec:training}.
The final valuation for that point in the search space was $\text{mean(EER)} + 1.96 \times \text{SD(EER)}$, using the combined measure of EER described in Section~\ref{sec:training} and aggregated across the 4~models.
Bayesian optimization sought to minimize this valuation.
The set of hyperparameters with the lowest valuation after all 31~search iterations was used during our final analyses.
Hyperparameters were tuned separately for each task and sampling rate.
The hyperparameter search space and the hyperparameters used for each model are all provided in the supplementary material.

\subsection{Final evaluation on test set}
\label{sec:evaluation}
There are two main scenarios for biometrics: authentication and identification.
In the authentication scenario, a user attempts to access a system by claiming to be a specific enrolled user and presenting their biometric sample (e.g., a fingerprint), and a decision is made based on the sample's similarity to the biometric template of that specific enrolled user.
In the identification scenario, a user attempts to access a system by claiming to be \textit{any} enrolled user and presenting their biometric sample, which is then compared against the biometric template of \textit{every} enrolled user to see if there is any match.
As more users are enrolled, it becomes increasingly difficult to detect impostors in the identification scenario, as the impostor's presented biometric sample need only be similar to any one enrolled user's biometric template.
This phenomenon has been formally studied using synthetic and real datasets~\cite{Friedman2020}.
Therefore, we consider only the authentication scenario, for which performance is expected to remain relatively consistent regardless of the number of enrolled users~\cite{Friedman2020}.

An important consideration for any biometric modality is how it compares to existing security methods.
For example, the 4-digit pin is one of the most common security methods for smartphones and is often used as a backup authentication method if biometrics fail.
If a biometric modality is less secure than a 4-digit pin, it would have little practical benefit on its own.
There are $10^4$ possible ways to construct a 4-digit pin with the numbers 0--9, so assuming each pin is equally likely to be chosen, there is a $10^{-4}$ chance that an impostor would correctly guess a specific user's pin.
Therefore, we provide measures of false rejection rate~(FRR) when false acceptance rate~(FAR) is fixed at $10^{-4}$, abbreviated FRR @ FAR $10^{-4}$.
According to the FIDO Biometrics Requirements~\cite{FIDO2020}, a biometric system should achieve a FRR @ FAR $10^{-4}$ of no more than 5\%.

We formed genuine and impostor pairs within the held-out test set for various test-retest intervals in a manner similar to the performance evaluation described in Section~\ref{sec:training}.
The enrollment set consists of the embeddings from the first $n$ non-overlapping subsequences of each recording in session~1 of R1.
Nine separate authentication sets were constructed using the embeddings from the first $n$ non-overlapping subsequences of each recording in session~2 of R1--9.
We then computed, for each authentication set separately, $S_{A,B}$ for all $A$ in the enrollment set and all $B$ in the authentication set using Equation~\ref{eq:sim}.

Using the above method, we have a maximum of 59~genuine pairs and 3422~impostor pairs (e.g., when authenticating with R1) and a minimum of 14~genuine pairs and 812~impostor pairs (when authenticating with R9).
This is fewer than the 10,000~impostor pairs needed to estimate FRR @ FAR $10^{-4}$.

\begin{figure}
    \centering
    \includegraphics[width=\linewidth]{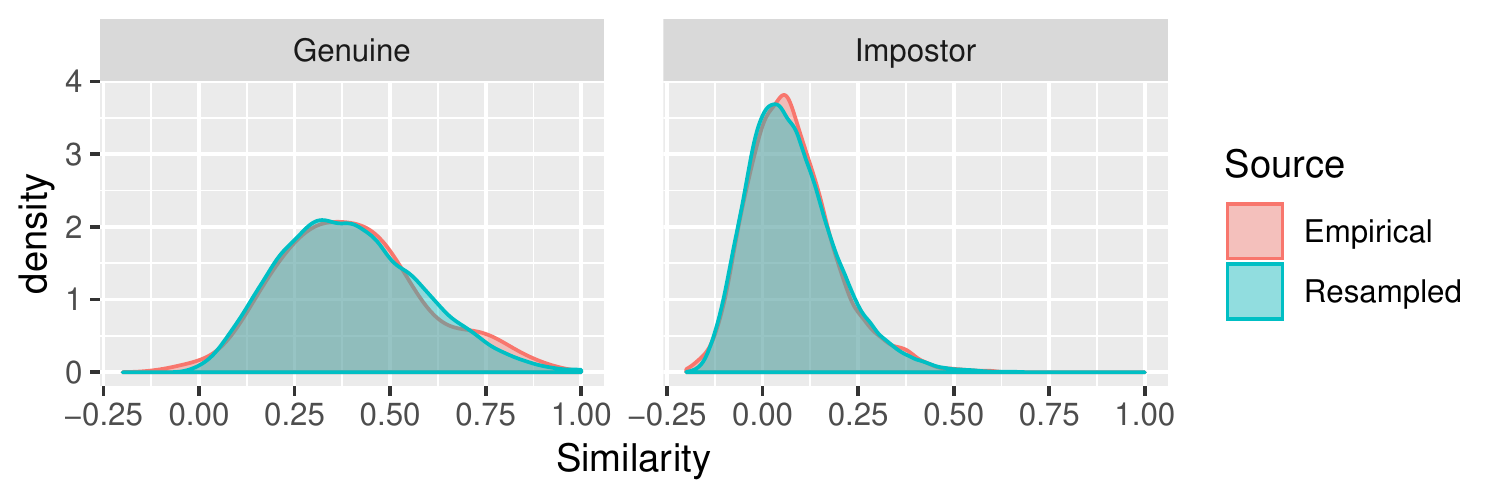}
    \caption{%
    A representative comparison between empirical and resampled similarity distributions.
    EERs based on the resampled distributions tend to be slightly pessimistic relative to EERs based on the empirical distributions.%
    }
    \label{fig:empirical_resampled}
\end{figure}

To enable the estimation of FRR @ FAR $10^{-4}$, we perform the following resampling approach.
We use the PearsonDS~\cite{Becker2017} R package to fit a Pearson family distribution to the empirical impostor similarity distribution for a given authentication set.
We sample 20,000~new impostor similarities from this fitted distribution and discard the empirical impostor similarities.
This fitted distribution closely matches the mean, variance, skewness, and kurtosis of our empirical distribution.
We do the same for the genuine similarity scores to balance the classes.
This provides us with enough data to be able to estimate FRR @ FAR $10^{-4}$.
By resampling from a fitted Pearson family distribution instead of, say, bootstrapping, we are able to sample values that are close to---but not exactly the same as---our empirical distribution of similarity scores.
See Figure~\ref{fig:empirical_resampled} for a representative comparison between empirical and resampled distributions.

We then construct a ROC curve and estimate EER for each authentication set separately, producing a total of 9~EER measures based on test-retest intervals as short as 20~minutes and as long as 37~months.

This entire process was repeated for each $n \in \{1, 5, 10\}$ to assess how authentication performance varies with the amount of data.

\subsection{Baseline models}
\subsubsection{Statistical baseline (STAR)}
One of the baselines we compare against is a statistical approach based on~\cite{Friedman2017} that we will refer to as STAR.
The same 4~folds and test set were used for STAR as were used for our metric learning approach.
Briefly, each recording was classified into fixations, saccades, post-saccadic oscillations~(PSOs), and noise using the modified Nystr\"{o}m and Holmqvist~(MNH) classification algorithm~\cite{Friedman2018}.
A set of over 1,000~features~\cite{Rigas2018} was extracted from the classified events.
Feature distributions were transformed to like-normal using the Box-Cox transformation, and only sufficiently normal features were kept.
Redundant features were removed next, either due to high correlations or due to multiple features measuring similar aspects (e.g., both mean and median of the same underlying feature).
Lastly, principal component analysis~(PCA) was performed, and the optimal set of features and principal components was determined through an iterative process.
We note that STAR was designed to use the full duration of each recording, so results for STAR are based on the full recording duration.

\subsubsection{DeepEyedentificationLive baseline (DEL)}
Our other baseline is the current state-of-the-art, DEL~\cite{Makowski2020}.
\rev{The code for DEL is publicly available\footnote{\rev{\url{https://osf.io/8es7z/}}}, and we reimplemented the network in PyTorch to keep all models under a single framework.}
\rev{Excluding classification layers, our implementation of DEL has 209,085,024~learnable parameters (69,052,160~in the slow subnet, 139,933,408~in the fast subnet, and 99,456~in the final set of layers after concatenation).}

To enable a more fair comparison against our own model, DEL underwent the same signal pre-processing, training, hyperparameter tuning, and test set evaluation as our own model, with only some minor differences.
We used an input length of 1024~time steps and the AdamW optimizer to match our model, instead of the original implementation's use of 1000~time steps and the Adam optimizer.
The input channels to the slow subnet were from Equations~\ref{eq:xslow}--\ref{eq:yslow}, and the input channels to the fast subnet were from Equations~\ref{eq:xfast}--\ref{eq:yfast} (we did not include stimulus position as an input, and we only had monocular gaze signals).
\rev{DEL was trained with PyTorch's \texttt{CrossEntropyLoss} instead of MS loss, and we trained the subnets individually before freezing their weights and training the final few fully-connected layers.}
\rev{After training, we removed the classification layer but still applied batch normalization and ReLU after the embedding layer, as is done in the public implementation.}
\rev{We did our best to match the weight initialization from Keras/Tensorflow (e.g., Keras uses a truncated normal distribution while PyTorch does not).}
Due to memory constraints, we used minibatches of 40~samples (with $k=4$ instead of 8) constructed in the manner described in Section~\ref{sec:training}.
Only 2~hyperparameters needed tuned for DEL: learning rate and weight decay.

\subsection{Hardware \& software}
\rev{All models (except the STAR and DEL baselines) were trained inside Docker containers on two Lambda Labs workstations.}
One workstation was equipped with dual NVIDIA GeForce RTX 2080 Ti GPUs (11~GB VRAM), an Intel i9-10920X CPU @ 3.50~GHz (12~cores), and 128~GB RAM.
The other workstation was equipped with dual NVIDIA GeForce RTX 3080 GPUs (10~GB VRAM), an Intel i9-10900X CPU @ 3.70~GHz (20~cores), and 64~GB RAM.
The Docker containers ran \rev{Ubuntu~18.04} and were set up with Python~3.7.10, PyTorch~1.9.0, and PML~\cite{Musgrave2020a} version~0.9.99.

\rev{Due to memory constraints, we needed to train the DEL baseline models on a different machine equipped with quad NVIDIA GeForce RTX A5000 GPUs (24~GB VRAM), an AMD Ryzen Threadripper PRO 3975WX CPU @ 3.5~GHz (32~cores), and 256~GB RAM.}
\rev{The slow and fast subnets were trained concurrently on separate GPUs to save time.}
\rev{DEL was trained inside a Docker container running Ubuntu~18.04 and set up with Python~3.7.11, PyTorch~1.10.0, and PML version~1.1.0.}

Each of our models took up to 1~hour to train for each fold on the RTX 3080 (24.0~training iterations per second), and up to 2~hours on the RTX 2080 Ti (13.3~training iterations per second).
\rev{The DEL baseline took an average of 2.8~hours to train for each fold on the RTX A5000, with the fast subnet often requiring more time to train than the slow subnet.}

STAR was run on a Windows~10 computer, equipped with an Intel i7-6700K CPU @ 4.00~GHz (4~cores) and 16~GB RAM.
The code was written in MATLAB~2020a and ran serially on the CPU.

Our full source code and trained models are available on the Texas State Digital Collections Repository at [TO BE PUBLISHED AFTER ACCEPTANCE].

\section{Results \& Discussion}
Due to prevalent usage of reading data in the eye movement biometrics literature, we use TEX @ 1000~Hz as our representative dataset.
The average performance measures of our approach on the held-out test set for TEX @ 1000~Hz are given in Table~\ref{tab:tex1000}, using each $n \in \{1, 5, 10\}$ (for Equation~\ref{eq:sim}) and each of R1--9 as the authentication set.
In Table~\ref{tab:other_tasks_rates}, we present results for all tasks and sampling rates using $n = 10$ and using R1 or R6 as the authentication set.
R1 has the shortest test-retest interval (approx. 20~minutes) and should, therefore, result in the best performance.
R6 has a test-retest interval of approx. 1~year, is exclusive to the test set, and contains all 59~unique test-set subjects.
Table~\ref{tab:baseline_results} shows results for the baseline models with TEX @ 1000~Hz, using R1 and R6 for authentication.
We note that the full duration of each recording was used for STAR, while only the first 10.240~s were used for DEL and our approach.

\rev{For a more comprehensive view of each model's performance, refer to the tables in the supplementary materials.}

\begin{table*}
    \centering
    \caption{%
        Results on the held-out test set for TEX @ 1000~Hz, varying the $n$ used for Equation~\ref{eq:sim} and the round used for the authentication set.
        Values are presented as mean (standard deviation) across the 4~models trained with 4-fold cross-validation.%
    }
    \label{tab:tex1000}
    \begin{tabular}{lcccccc}
        \toprule
        \multirow{2}[2]{*}{$n$} & \multirow{2}[2]{*}{Round} & \multirow{2}[2]{*}{EER} & \multicolumn{4}{c}{FRR @ FAR} \\\cmidrule{4-7}
        {} & {} & {} & $10^{-1}$ & $10^{-2}$ & $10^{-3}$ & $10^{-4}$ \\
        \midrule
        \multirow{9}{*}{1} & 1 & 0.2237 (0.0115) & 0.3866 (0.0256) & 0.7483 (0.0377) & 0.9370 (0.0378) & 0.9870 (0.0168) \\
        {} & 2 & 0.2903 (0.0214) & 0.5415 (0.0402) & 0.8584 (0.0295) & 0.9673 (0.0183) & 0.9947 (0.0044) \\
        {} & 3 & 0.3045 (0.0214) & 0.5582 (0.0339) & 0.8684 (0.0121) & 0.9775 (0.0082) & 0.9979 (0.0032) \\
        {} & 4 & 0.3103 (0.0170) & 0.5754 (0.0334) & 0.8697 (0.0241) & 0.9714 (0.0128) & 0.9959 (0.0028) \\
        {} & 5 & 0.2760 (0.0143) & 0.4741 (0.0120) & 0.7965 (0.0224) & 0.9738 (0.0159) & 0.9977 (0.0040) \\
        {} & 6 & 0.2783 (0.0249) & 0.5134 (0.0288) & 0.8434 (0.0303) & 0.9640 (0.0204) & 0.9939 (0.0078) \\
        {} & 7 & 0.3051 (0.0260) & 0.5349 (0.0521) & 0.8272 (0.0382) & 0.9549 (0.0218) & 0.9904 (0.0068) \\
        {} & 8 & 0.3036 (0.0253) & 0.5502 (0.0529) & 0.8765 (0.0317) & 0.9942 (0.0101) & 0.9995 (0.0009) \\
        {} & 9 & 0.2841 (0.0361) & 0.5623 (0.0427) & 0.9027 (0.0617) & 0.9787 (0.0191) & 0.9962 (0.0042) \\\cmidrule{2-7}
        \multirow{9}{*}{5} & 1 & 0.1488 (0.0075) & 0.2206 (0.0193) & 0.6284 (0.0579) & 0.8790 (0.0543) & 0.9619 (0.0240) \\
        {} & 2 & 0.1952 (0.0100) & 0.3449 (0.0188) & 0.7801 (0.0429) & 0.9577 (0.0237) & 0.9955 (0.0039) \\
        {} & 3 & 0.2260 (0.0154) & 0.3981 (0.0290) & 0.7709 (0.0284) & 0.9318 (0.0292) & 0.9827 (0.0111) \\
        {} & 4 & 0.2008 (0.0144) & 0.3694 (0.0266) & 0.7831 (0.0211) & 0.9492 (0.0191) & 0.9920 (0.0042) \\
        {} & 5 & 0.2037 (0.0144) & 0.3638 (0.0335) & 0.7451 (0.0089) & 0.9162 (0.0238) & 0.9679 (0.0213) \\
        {} & 6 & 0.2178 (0.0105) & 0.4058 (0.0259) & 0.8030 (0.0389) & 0.9608 (0.0233) & 0.9895 (0.0060) \\
        {} & 7 & 0.2367 (0.0166) & 0.4440 (0.0276) & 0.8070 (0.0251) & 0.9455 (0.0286) & 0.9792 (0.0186) \\
        {} & 8 & 0.2432 (0.0285) & 0.4433 (0.0446) & 0.8192 (0.0434) & 0.9679 (0.0179) & 0.9988 (0.0018) \\
        {} & 9 & 0.1815 (0.0125) & 0.3537 (0.0732) & 0.8955 (0.0762) & 0.9793 (0.0215) & 0.9939 (0.0070) \\\cmidrule{2-7}
        \multirow{9}{*}{10} & 1 & 0.1420 (0.0032) & 0.2038 (0.0084) & 0.6272 (0.0646) & 0.8758 (0.0524) & 0.9665 (0.0200) \\
        {} & 2 & 0.1924 (0.0103) & 0.3426 (0.0184) & 0.7782 (0.0437) & 0.9464 (0.0263) & 0.9865 (0.0080) \\
        {} & 3 & 0.2110 (0.0147) & 0.3847 (0.0392) & 0.7735 (0.0425) & 0.9371 (0.0295) & 0.9829 (0.0158) \\
        {} & 4 & 0.1952 (0.0133) & 0.3397 (0.0342) & 0.7530 (0.0142) & 0.9254 (0.0213) & 0.9802 (0.0167) \\
        {} & 5 & 0.1889 (0.0112) & 0.3354 (0.0368) & 0.7073 (0.0205) & 0.8899 (0.0301) & 0.9490 (0.0336) \\
        {} & 6 & 0.2110 (0.0148) & 0.3915 (0.0364) & 0.7900 (0.0438) & 0.9451 (0.0294) & 0.9845 (0.0111) \\
        {} & 7 & 0.2277 (0.0140) & 0.4263 (0.0286) & 0.7834 (0.0335) & 0.9315 (0.0271) & 0.9829 (0.0133) \\
        {} & 8 & 0.2156 (0.0169) & 0.4217 (0.0340) & 0.8586 (0.0399) & 0.9803 (0.0102) & 0.9983 (0.0019) \\
        {} & 9 & 0.2017 (0.0035) & 0.3798 (0.0391) & 0.8766 (0.0866) & 0.9716 (0.0280) & 0.9922 (0.0077) \\
        \bottomrule
    \end{tabular}
\end{table*}
\begin{table*}
    \centering
    \caption{%
        Results on the held-out test set for every task and sampling rate.
        For brevity, results are only shown when using R1 and R6 for the authentication set, and only when using $n=10$ in Equation~\ref{eq:sim}.
        Values are presented as mean (standard deviation) across the 4~models trained with 4-fold cross-validation.
        The best result for each round is bolded.\\
        *: For BLG, 3~subjects (1, 120, and 180) were excluded at test time for having a recording with a duration less than 10.240~s.%
    }
    \label{tab:other_tasks_rates}
    \begin{adjustbox}{width=\textwidth}
    \begin{tabular}{lcccccc}
        \toprule
        \multirow{2}[2]{*}{Task @ Rate} & \multirow{2}[2]{*}{Round} & \multirow{2}[2]{*}{EER} & \multicolumn{4}{c}{FRR @ FAR} \\\cmidrule{4-7}
        {} & {} & {} & $10^{-1}$ & $10^{-2}$ & $10^{-3}$ & $10^{-4}$ \\
        
        \midrule
        \multirow{2}{*}{HSS @ 1000~Hz} & 1 & \textbf{0.1125 (0.0071)} & \textbf{0.1273 (0.0154)} & \textbf{0.4867 (0.0299)} & 0.8010 (0.0891) & 0.9343 (0.0542) \\
        {} & 6 & 0.2310 (0.0082) & 0.4630 (0.0174) & 0.8377 (0.0302) & 0.9495 (0.0210) & 0.9808 (0.0096) \\\cmidrule{2-7}
        \multirow{2}{*}{VD1 @ 1000~Hz} & 1 & 0.1694 (0.0112) & 0.2770 (0.0276) & 0.7343 (0.0460) & 0.9288 (0.0498) & 0.9789 (0.0298) \\
        {} & 6 & 0.3151 (0.0179) & 0.6099 (0.0297) & 0.9125 (0.0016) & 0.9901 (0.0080) & 0.9979 (0.0030) \\\cmidrule{2-7}
        \multirow{2}{*}{FXS @ 1000~Hz} & 1 & 0.2136 (0.0102) & 0.4260 (0.0389) & 0.8664 (0.0329) & 0.9763 (0.0205) & 0.9953 (0.0064) \\
        {} & 6 & 0.3654 (0.0107) & 0.7113 (0.0218) & 0.9518 (0.0040) & 0.9949 (0.0021) & 0.9994 (0.0005) \\\cmidrule{2-7}
        \multirow{2}{*}{RAN @ 1000~Hz} & 1 & 0.1459 (0.0079) & 0.2293 (0.0210) & 0.7602 (0.0619) & 0.9481 (0.0445) & 0.9836 (0.0175) \\
        {} & 6 & 0.2626 (0.0194) & 0.5294 (0.0385) & 0.9188 (0.0160) & 0.9927 (0.0081) & 0.9987 (0.0017) \\\cmidrule{2-7}
        \multirow{2}{*}{*BLG @ 1000~Hz} & 1 & 0.1404 (0.0112) & 0.2140 (0.0273) & 0.6835 (0.0125) & 0.8902 (0.0355) & 0.9488 (0.0272) \\
        {} & 6 & 0.2833 (0.0132) & 0.5578 (0.0313) & 0.8490 (0.0181) & 0.9315 (0.0282) & 0.9528 (0.0330) \\ 
        
        \midrule
        \multirow{2}{*}{TEX @ 1000~Hz} & 1 & 0.1420 (0.0032) & 0.2038 (0.0084) & 0.6272 (0.0646) & 0.8758 (0.0524) & 0.9665 (0.0200) \\
        {} & 6 & \textbf{0.2110 (0.0148)} & \textbf{0.3915 (0.0364)} & 0.7900 (0.0438) & 0.9451 (0.0294) & 0.9845 (0.0111) \\ 
        
        \midrule
        \multirow{2}{*}{TEX @ 500~Hz} & 1 & 0.1667 (0.0155) & 0.3110 (0.0577) & 0.8315 (0.0293) & 0.9928 (0.0051) & 1.0000 (0.0000) \\
        {} & 6 & 0.2764 (0.0158) & 0.5618 (0.0317) & 0.9280 (0.0179) & 0.9975 (0.0020) & 0.9999 (0.0001) \\\cmidrule{2-7}
        \multirow{2}{*}{TEX @ 250~Hz} & 1 & 0.1661 (0.0091) & 0.2540 (0.0201) & 0.6916 (0.0439) & 0.9597 (0.0399) & 0.9999 (0.0001) \\
        {} & 6 & 0.2660 (0.0157) & 0.4845 (0.0280) & 0.8560 (0.0304) & 0.9790 (0.0196) & 0.9959 (0.0067) \\\cmidrule{2-7}
        \multirow{2}{*}{TEX @ 125~Hz} & 1 & 0.1875 (0.0106) & 0.3123 (0.0281) & 0.6327 (0.0277) & 0.7990 (0.0276) & 0.8642 (0.0289) \\
        {} & 6 & 0.2499 (0.0165) & 0.4780 (0.0359) & 0.7994 (0.0335) & 0.9159 (0.0219) & 0.9504 (0.0201) \\\cmidrule{2-7}
        \multirow{2}{*}{TEX @ 50~Hz} & 1 & 0.2371 (0.0436) & 0.4152 (0.1301) & 0.5411 (0.1669) & \textbf{0.5558 (0.1720)} & \textbf{0.5577 (0.1730)} \\
        {} & 6 & 0.2466 (0.0394) & 0.4742 (0.1329) & 0.7981 (0.1169) & \textbf{0.8150 (0.1075)} & \textbf{0.8173 (0.1064)} \\\cmidrule{2-7}
        \multirow{2}{*}{TEX @ 31.25~Hz} & 1 & 0.2666 (0.0246) & 0.5147 (0.0628) & 0.7356 (0.0476) & 0.7881 (0.0645) & 0.8079 (0.0779) \\
        {} & 6 & 0.3047 (0.0281) & 0.5902 (0.0591) & \textbf{0.7796 (0.0368)} & 0.8282 (0.0537) & 0.8404 (0.0600) \\
        \bottomrule
    \end{tabular}
    \end{adjustbox}
\end{table*}
\begin{table*}
    \centering
    \caption{%
        Baseline performance measures computed on the held-out test set using TEX @ 1000~Hz.
        For brevity, results are only shown when using R1 and R6 for the authentication set, and only when using $n=10$ in Equation~\ref{eq:sim} (full recording duration used for STAR).
        Values are given as mean (standard deviation), aggregated across 4~models trained via 4-fold cross-validation.
        The best result for each round is bolded.%
    }
    \label{tab:baseline_results}
    
    \begin{adjustbox}{width=\textwidth}
    \rev{
    \begin{tabular}{lcccccc}
        \toprule
        \multirow{2}[2]{*}{Model} & \multirow{2}[2]{*}{Round} & \multirow{2}[2]{*}{EER} & \multicolumn{4}{c}{FRR @ FAR} \\\cmidrule{4-7}
        {} & {} & {} & $10^{-1}$ & $10^{-2}$ & $10^{-3}$ & $10^{-4}$ \\
        \midrule
        \multirow{2}{*}{STAR} & 1 & 0.1563 (0.0487) & 0.2249 (0.1101) & 0.6240 (0.2479) & \textbf{0.8040 (0.1962)} & 0.9107 (0.0917) \\
        {} & 6 & 0.2461 (0.0620) & 0.4161 (0.1188) & 0.8284 (0.1727) & 0.9237 (0.0765) & 0.9727 (0.0281) \\\cmidrule{2-7}
        \multirow{2}{*}{DEL} & 1 & 0.4295 (0.0213) & 0.8001 (0.0240) & 0.9706 (0.0165) & 0.9955 (0.0033) & 0.9992 (0.0005) \\
        {} & 6 & 0.4748 (0.0197) & 0.8465 (0.0301) & 0.9867 (0.0172) & 0.9973 (0.0034) & 0.9991 (0.0010) \\ \cmidrule{2-7}
        \multirow{2}{*}{DEL (slow)} & 1 & 0.1559 (0.0394) & 0.2226 (0.0854) & \textbf{0.5822 (0.1083)} & 0.8558 (0.1072) & 0.9550 (0.0483) \\
        {} & 6 & 0.2314 (0.0336) & 0.4053 (0.0807) & \textbf{0.7619 (0.0662)} & 0.9241 (0.0438) & 0.9687 (0.0230) \\ \cmidrule{2-7}
        \multirow{2}{*}{DEL (fast)} & 1 & 0.2111 (0.0377) & 0.3631 (0.0737) & 0.6995 (0.0371) & 0.8208 (0.0766) & \textbf{0.8691 (0.0973)} \\
        {} & 6 & 0.2673 (0.0326) & 0.5296 (0.0612) & 0.8208 (0.0189) & \textbf{0.8967 (0.0317)} & \textbf{0.9146 (0.0403)} \\
        \midrule
        \multirow{2}{*}{Ours} & 1 & \textbf{0.1420 (0.0032)} & \textbf{0.2038 (0.0084)} & 0.6272 (0.0646) & 0.8758 (0.0524) & 0.9665 (0.0200) \\
        {} & 6 & \textbf{0.2110 (0.0148)} & \textbf{0.3915 (0.0364)} & 0.7900 (0.0438) & 0.9451 (0.0294) & 0.9845 (0.0111) \\
        \bottomrule
    \end{tabular}
    }
    \end{adjustbox}
\end{table*}

\begin{figure}
    \centering
    
    \subfloat[][TEX @ 1000~Hz]{
        \label{fig:hist-tex1000}
        \includegraphics[height=0.13\textheight]{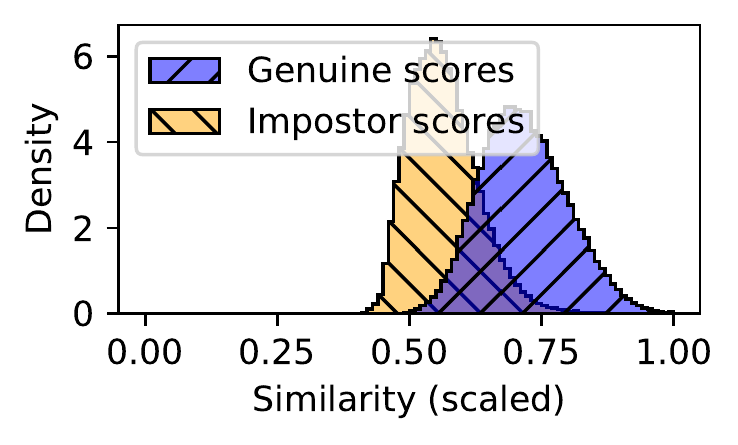}
    }
    \subfloat[][STAR]{
        \label{fig:hist-star}
        \includegraphics[height=0.13\textheight]{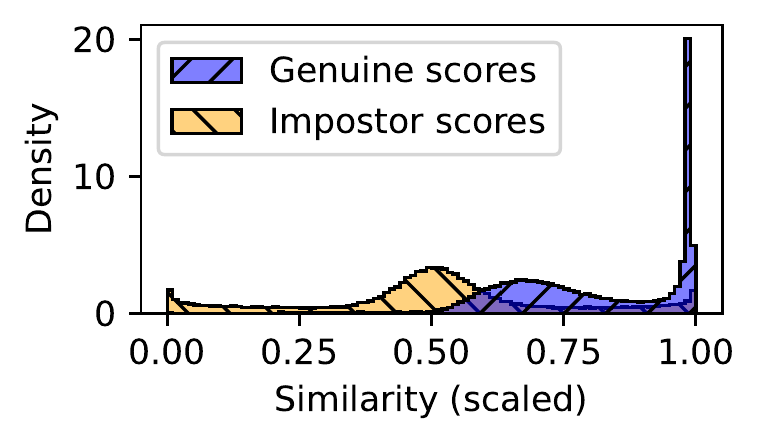}
    }
    \rev{
    \subfloat[][DEL]{
        \label{fig:hist-del}
        \includegraphics[height=0.13\textheight]{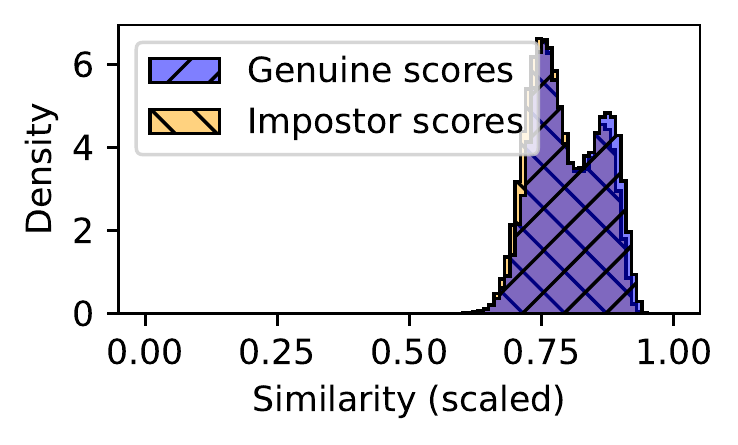}
    }
    }
    
    \subfloat[][HSS]{
        \label{fig:hist-hss1000}
        \includegraphics[height=0.13\textheight]{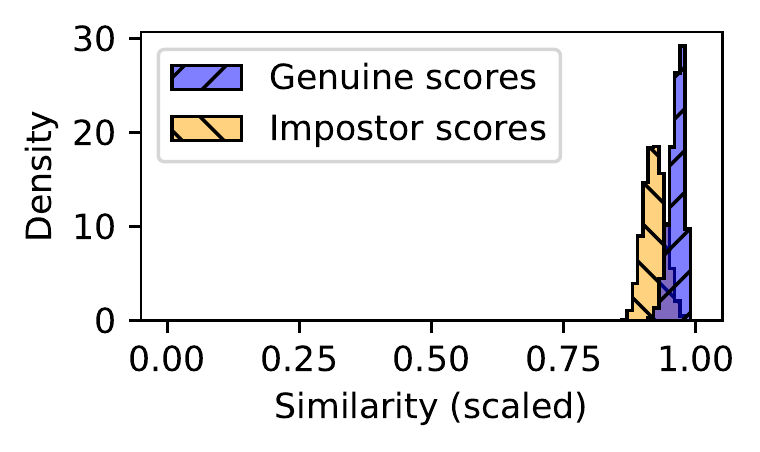}
    }
    \subfloat[][VD1]{
        \label{fig:hist-vd11000}
        \includegraphics[height=0.13\textheight]{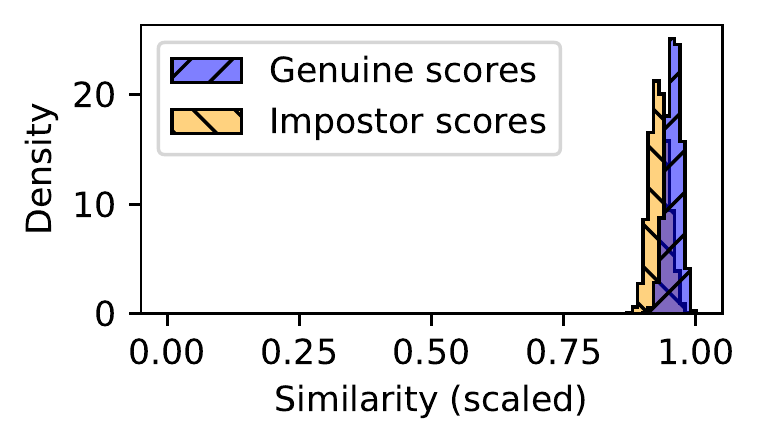}
    }
    \subfloat[][FXS]{
        \label{fig:hist-fxs1000}
        \includegraphics[height=0.13\textheight]{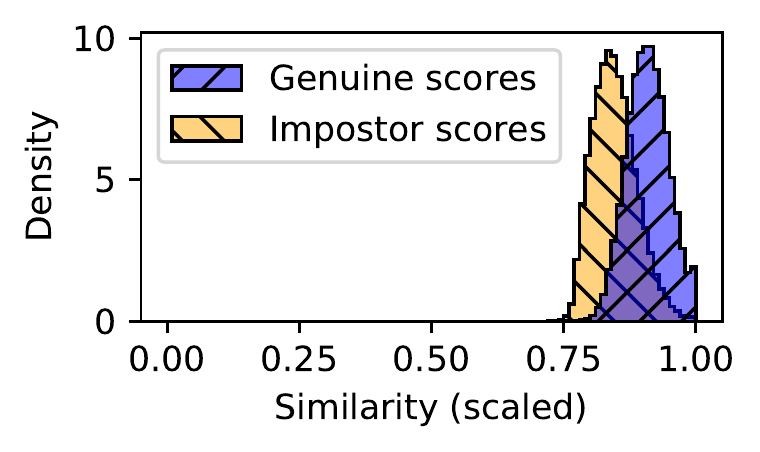}
    }
    
    \subfloat[][RAN]{
        \label{fig:hist-ran1000}
        \includegraphics[height=0.13\textheight]{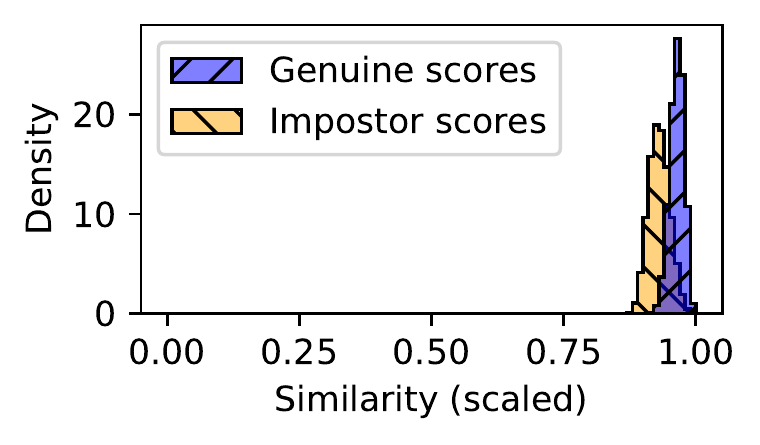}
    }
    \subfloat[][BLG]{
        \label{fig:hist-blg1000}
        \includegraphics[height=0.13\textheight]{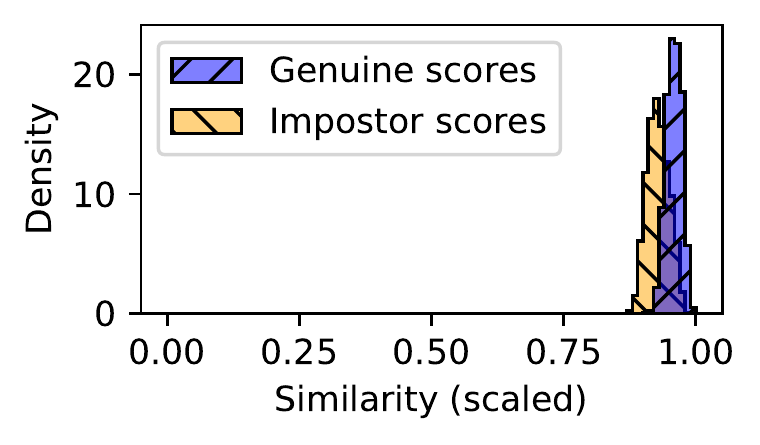}
    }
    \subfloat[][TEX @ 500~Hz]{
        \label{fig:hist-tex0500}
        \includegraphics[height=0.13\textheight]{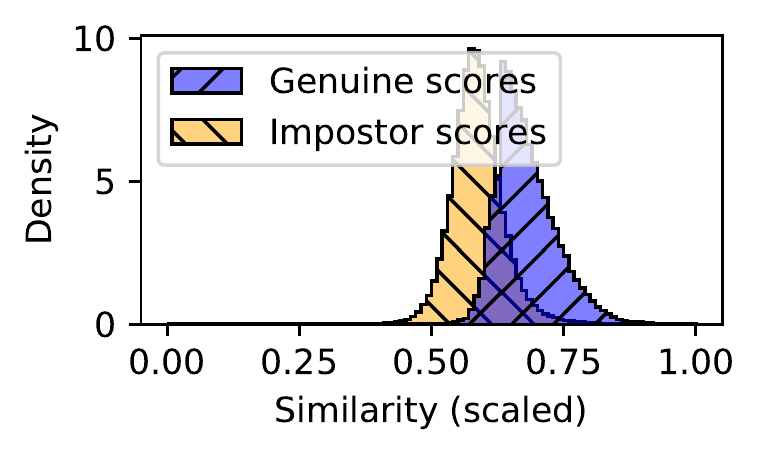}
    }
    
    \subfloat[][TEX @ 250~Hz]{
        \label{fig:hist-tex0250}
        \includegraphics[height=0.13\textheight]{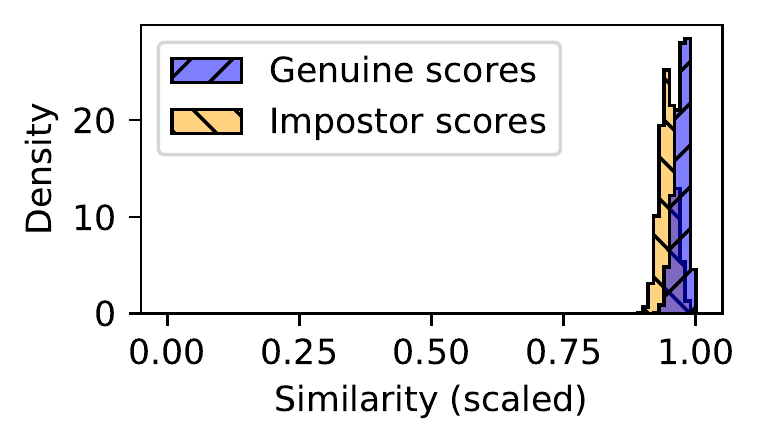}
    }
    \subfloat[][TEX @ 125~Hz]{
        \label{fig:hist-tex0125}
        \includegraphics[height=0.13\textheight]{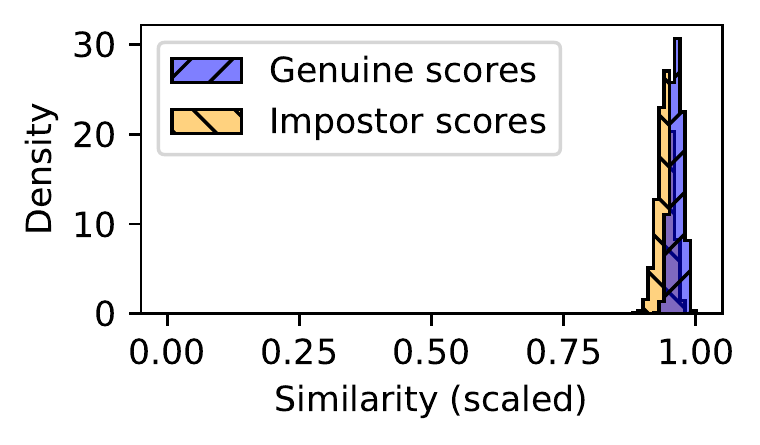}
    }
    \subfloat[][TEX @ 50~Hz]{
        \label{fig:hist-tex0050}
        \includegraphics[height=0.13\textheight]{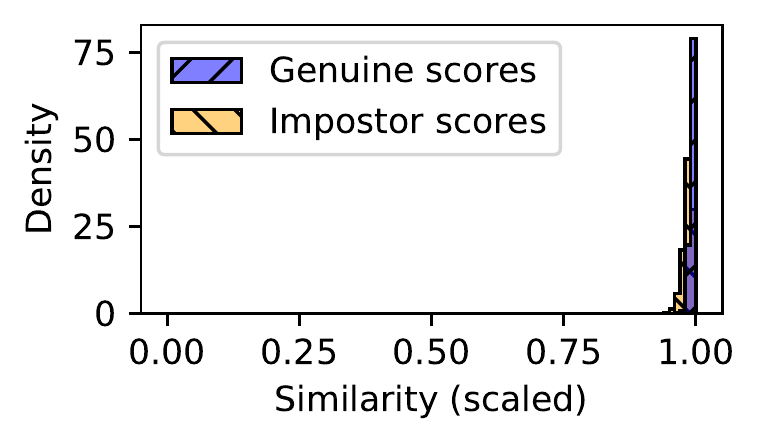}
    }
    
    \subfloat[][TEX @ 31.25~Hz]{
        \label{fig:hist-tex0031}
        \includegraphics[height=0.13\textheight]{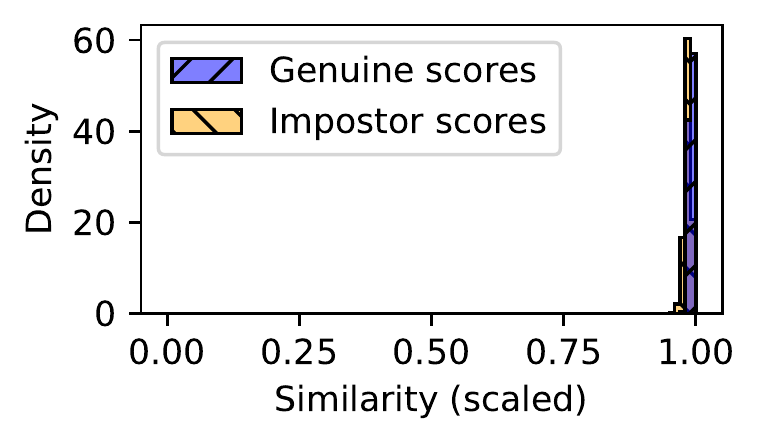}
    }
    \caption[Similarity distributions for genuine and impostor pairs.]{Plots of the similarity distributions for genuine and impostor pairs.
    Each plot contains the similarities on the held-out test set computed separately for each of the 4~models trained with 4-fold cross-validation, using $n = 10$ for Equation~\ref{eq:sim} and R1 for authentication.
    Since cosine similarity is bounded from~-1 to~1, we scaled the similarities to lie between~0 and~1 before plotting.
    A bin width of~0.01 was used, and the area under each curve sums to~1.}
    \label{fig:all-hists}
\end{figure}

\begin{figure}
    \centering
    \subfloat[][Varying tasks]{
        \label{fig:roc-alltasks}
        \includegraphics[width=0.45\linewidth]{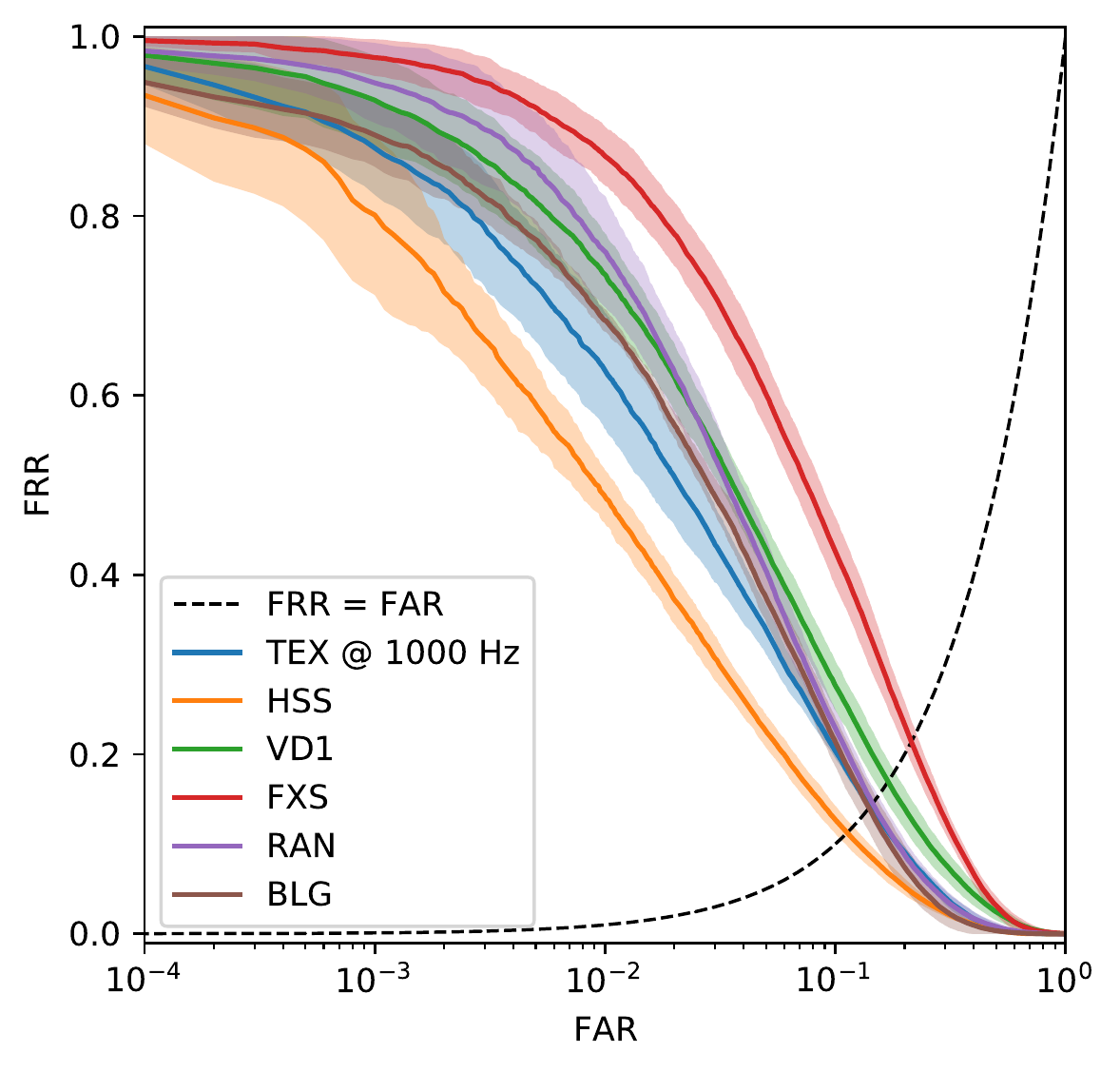}
    }
    \subfloat[][Varying sampling rate]{
        \label{fig:roc-allrates}
        \includegraphics[width=0.45\linewidth]{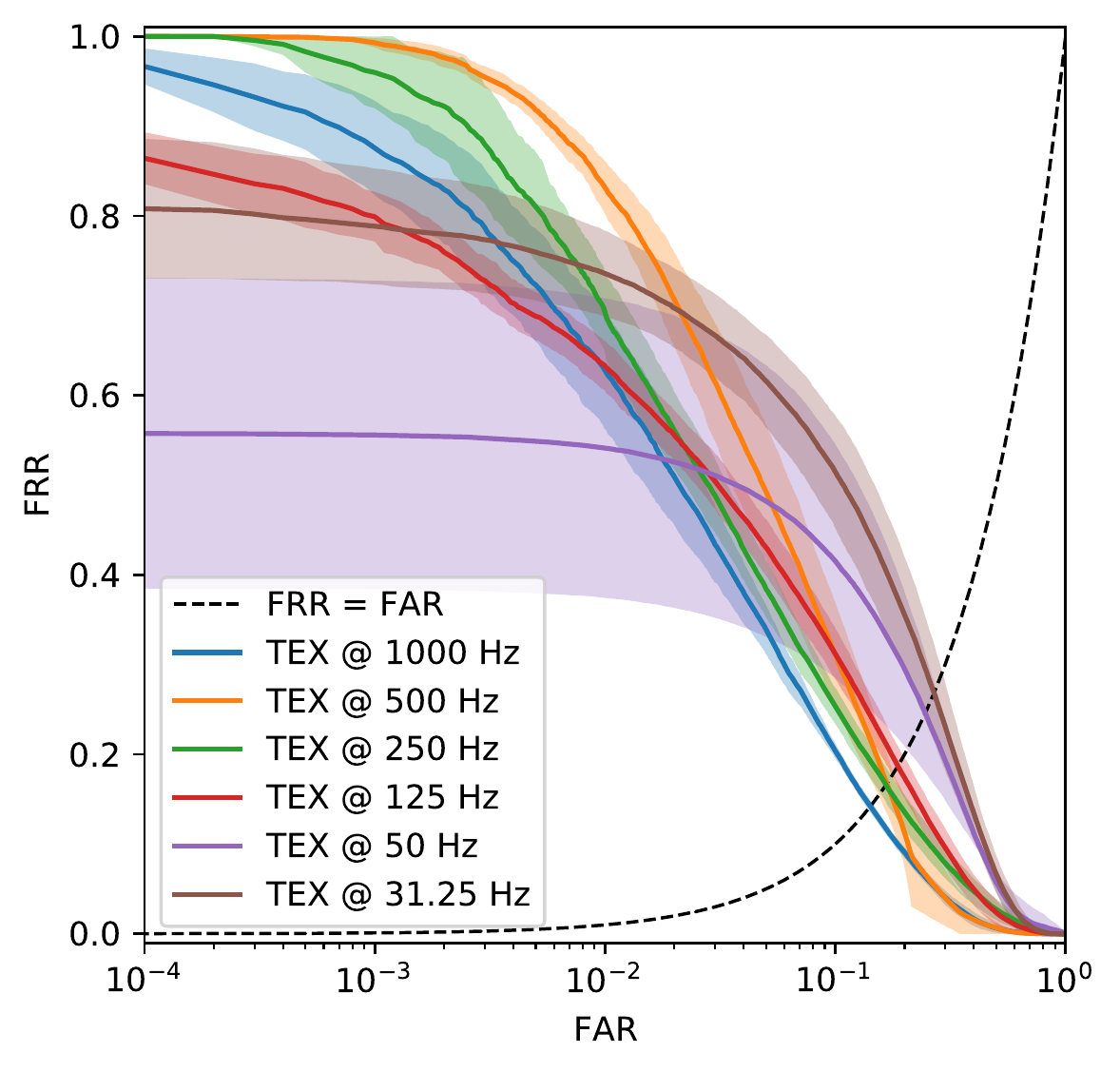}
    }\\
    \rev{
    \subfloat[][Baselines]{
        \label{fig:roc-baselines}
        \includegraphics[width=0.45\linewidth]{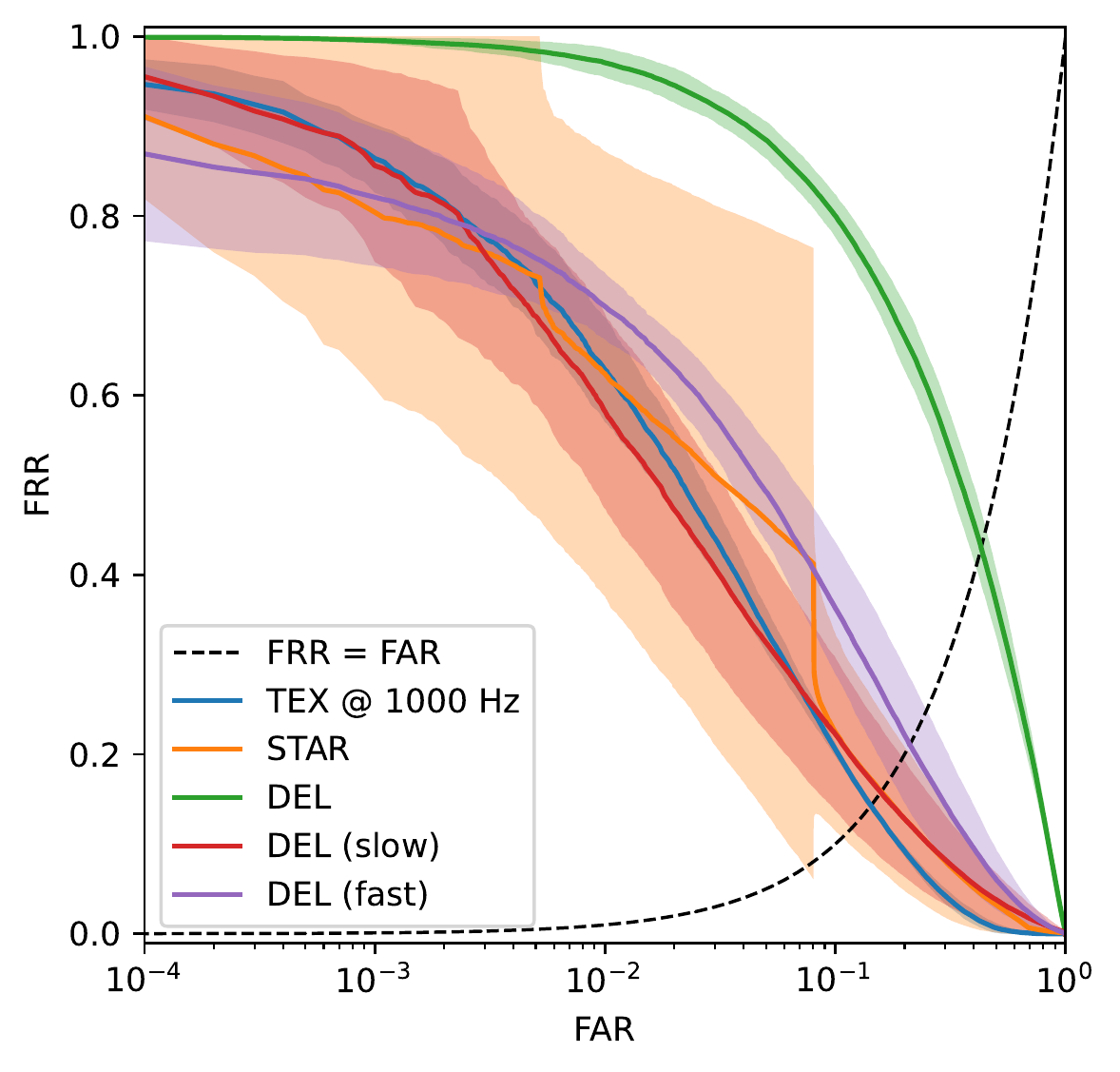}
    }
    }
    \caption[ROC curves for each model.]{%
    ROC curves to provide a qualitative assessment of model performance using $n = 10$ and R1 for authentication.
    The horizontal axis is log-scaled false acceptance rate~(FAR).
    The vertical axis is false rejection rate~(FRR).
    Each ROC curve represents the mean performance across 4~models trained with 4-fold cross-validation, and each shaded region indicates $\pm$1~SD about the mean.
    The point where the dashed line intersects each ROC curve indicates the EER for that curve.%
    }
    \label{fig:all-rocs}
\end{figure}

In addition to the quantitative results in the aforementioned tables, we also present some qualitative results.
For these figures, each model uses $n = 10$ (except STAR, which uses the full recording duration) and R1 for authentication.
Figure~\ref{fig:all-hists} shows a comparison between the genuine and impostor similarity score distributions.
An ROC curve for each model is presented in Figure~\ref{fig:all-rocs}.


\subsection{Comparison to baselines}
\subsubsection{STAR}
Looking at Table~\ref{tab:baseline_results}, compared to STAR, our approach results in lower EERs and is more stable across folds.

The genuine vs impostor distributions (Figure~\ref{fig:hist-star}) and the ROC curve (Figure~\ref{fig:roc-baselines}) for STAR deserve additional discussion.
The genuine and impostor distributions are not unimodal in the presented figure, but this is because the figure includes 4~separate genuine/impostor distributions (one per model).
The mean ROC curve shows a sharp increase in FRR when FAR is approximately 0.08, and another around 0.005~FAR.
We believe these abnormalities are due to different features and principal components being included for each fold, resulting in vastly different performance between models.
Perhaps it would have been better to use a consistent set of features and principal components across folds.

STAR requires event classification and manual feature extraction, making it much harder to employ in different datasets and for different tasks than our end-to-end approach.
It was also designed to use the full duration of each recording, limiting its use in practice.

Regarding the discrepancy between our results with STAR and the original results from~\cite{Friedman2017} (where an EER of 10\% was achieved on data separated by approximately 1~year), there are many contributing factors.

First, our EER estimates are based on different data than the original study.
We evaluated with as many as 59~subjects, did not remove signal after the end of reading, and did not exclude any subjects from the analysis regardless of data quality.
In contrast, the original study used either 149~subjects (SBA-ST) or 34~subjects (SBA-LT), removed signal after the end of reading, and screened subjects with ``low recording quality'' or who had ``excessively noisy recordings.''

Second, we made several changes to the original approach: we simplified the normality transformations by always using Box-Cox instead of trying several different standard transformation functions; we did not winsorize the distributions to try to improve normality; we tested normality by checking skewness and kurtosis instead of using the chi-square test; we measured reliability on a set of data disjoint from our test set; and we determined the best set of features and number of principal components using mean EER across rounds rather than rank-1 identification rate.

Third, we did not have access to several oculomotor plant characteristic~(OPC) features that were present in the original study, some of which were found to be highly reliable in the original study.

\subsubsection{DEL}
\rev{We were unable to get reasonable performance with the full DEL model.}
\rev{Since the fast and slow subnets both performed well on their own, we included their performance in our results.}
\rev{Looking at Table~\ref{tab:baseline_results}, we see that the full DEL model had near-random performance with 42.95\% R1 EER.}
\rev{The slow and fast subnets individually performed better than the full DEL model, achieving R1 EERs of 15.59\% and 21.11\%, respectively.}
\rev{Our proposed model outperforms the DEL baseline, though the difference between the slow subnet and our model may not be statistically significant.}

\rev{As with STAR, the genuine and impostor distributions~(Fig.~\ref{fig:hist-del}) are not unimodal, but this is because the figure includes 4~separate genuine/impostor distributions (one per model).}

\rev{We note that our approach outperforms DEL, despite ours having 440x fewer learnable parameters (145x fewer than the slow subnet).}
This significant reduction in model complexity may enable more power-efficient implementations in certain target settings, such as embedded environments.

Regarding the discrepancy between our results with DEL and the original results from~\cite{Makowski2020} (where an EER @ 10~s of 5\% was achieved on data separated by $\ge$1--4~weeks), we note that we \rev{used a different data set and} evaluated the model on a held-out test set of as many as 59~subjects (compared to 25~subjects used in the original study).
Additionally, we had only 1~enrolled recording per subject in the test set, whereas the original study had multiple.
We also determined similarity differently.
The original DEL study~\cite{Makowski2020} checks if \textit{any} window from \textit{any} enrolled sequence is sufficiently similar to \textit{any} window from the presented sequence during authentication.
In contrast, we measured the mean similarity across each pair of temporally-aligned windows from the enrolled sequence and the presented sequence during authentication.

\subsection{Authentication accuracy vs test-retest interval}
Looking at Table~\ref{tab:tex1000}, we see that performance when authenticating on R1 (test-retest interval of approx. 20~minutes) is significantly better than later rounds.
This matches our expectation.

We note that R1--6 all have 59~subjects, R7 has 35, R8 has 31, and R9 has 14.
The reduction in subject count for R7--9 may partially explain the reduction in EER despite the increase in test-retest interval.
We also note that data for other subjects from R1--5 were present in the training and validation set, while R6--9 were exclusive to the test set.

\subsection{Authentication accuracy vs recording duration}
Looking at Table~\ref{tab:tex1000}, we see that our model still performs better than chance (50\% EER) using just the first 1.024~s ($n = 1$) from each recording for both enrollment and authentication.
Performance drastically improves when the first 5.120~s ($n = 5$) are used but does not improve much further when using the first 10.240~s ($n = 10$).

\rev{Although we focused on low data requirements, it may be worth mentioning that in an additional analysis, we evaluated the TEX @ 1000~Hz model using (roughly) the full recording duration ($n = 58$, or 59.392~s) for both enrollment and authentication.}
\rev{In this higher data setting, our model achieved a mean R1 EER of 10.52\%---an improvement of nearly 4~percentage points compared to using $n=10$.}

\subsection{Authentication accuracy vs sampling rate}
Looking at the bottom half of Table~\ref{tab:other_tasks_rates}, we see that R1 EER worsened as sampling rate was reduced, which aligns with our expectations.
However, R6 EER monotonically improved (slightly) as sampling rate was degraded, starting from 500~Hz down to as low as 50~Hz.
Figure~\ref{fig:all-hists} shows that as sampling rate is degraded and fewer time steps are present in the input, the models gradually become less capable of producing embeddings that are highly dissimilar.

Interestingly, R1 FRR @ FAR $10^{-4}$ was its lowest at 50~Hz (see Table~\ref{tab:other_tasks_rates}).
Of course, a FRR of 56\% is still unusable in practice, and the variance across folds was largest for 50~Hz; but this was an interesting result nonetheless.

\subsection{Authentication accuracy vs task}
Looking at the top half of Table~\ref{tab:other_tasks_rates}, we find that HSS, a low-cognitive-load task, resulted in the best R1 performance across tasks.
TEX resulted in the best R6 performance across tasks, but HSS was still competitive despite requiring significantly less mental effort from the participants.
Unsurprisingly, FXS resulted in the worst performance of all the tasks, but still did better than chance (50\% EER) even on R6.

\rev{\subsection{Explanation of high error rates}}
\rev{The results presented herein are nominally worse than those of many prior works in the literature.}
\rev{We highlight, for the following reasons, that this relatively poor performance is due to our work attempting to solve a harder problem that is more practically relevant and thus would be more indicative of real-world performance with the selected architecture.}
\rev{Within this more realistic scenario, our proposed architecture outperformed all state-of-the-art models.}

\rev{\textbf{Low data setting.}\quad Our approach for authentication requires very little data (up to 10.240~s collected during one sitting), during both enrollment and verification.}
\rev{The majority of prior works require significantly more data.}
\rev{For example, DEL~\cite{Makowski2020} uses 9~separate trials (totaling 26.250~s) collected over a period of at least 3~weeks for enrollment.}
\rev{The statistical method by Friedman et al.~\cite{Friedman2017} uses 60~s of data during both enrollment and verification.}
\rev{We believe that it is important for future studies to focus on requiring less data to improve the practical utility of eye movements as a biometric.}

\rev{\textbf{No data screening.}\quad We did not clean the data set at all (beyond any prior screening employed for the GazeBase data set itself).}
\rev{As a result, the data we used included noisy signals littered with NaN values, and some windows of data used during training and evaluation had a significant amount of missing data.}
\rev{For an extreme example, the test set for TEX @ 1000~Hz contains 23 recordings with 100\% NaN values in the first 10~windows (10.240~s).}
\rev{Such recordings could not possibly be reasonably classified, likely contributing to the very high values for FRR @ FAR $10^{-4}$.}

\rev{\textbf{Held-out test set.}\quad We used a separate held-out set of data for the final evaluation of our model.}
\rev{This held-out set contains nearly 50\% of all recordings in GazeBase, resulting in much less data available for training.}
\rev{It also contains data from 59~subjects which is a larger population than many prior studies consider.}
\rev{Most prior studies do not use a separate held-out set of data, so their estimates of model performance may be more biased (in their favor).}

\rev{\textbf{Pessimistic resampling.}\quad Our approach of resampling the similarity scores with a Pearson family distribution tended to produce pessimistic estimates of model performance that are likely more indicative of real-world performance.}
\rev{While our measures of FRR @ FAR $10^{-4}$ are very high to the point of limited practical use, we note that our study is the first in the field to report measures of FRR @ FAR $10^{-4}$.}

\subsection{Limitations}
There is an implicit assumption that the embeddings of each subsequence within a recording come from the same distribution.
This assumption is necessary for the metric learning model to learn a well-clustered embedding space, as the subsequence embeddings for a given subject should follow some central tendency.
However, during TEX, for instance, this assumption is almost certainly violated for whichever subsequence inevitably contains the large return saccade that occurs when a participant finishes reading the text and starts re-reading it.
It may be advantageous to exclude such anomalous subsequences during training.

We note that degrading sampling rate alone is not sufficient to emulate other eye trackers with worse signal quality.
There would also be differences in other eye tracking signal quality metrics including spatial accuracy, spatial precision, temporal precision, linearity, and crosstalk.

The EERs presented in the present study (and virtually every other study in the field) are based on a threshold determined on the test set data.
We note that doing so essentially leaks test set data into the decision process.
Of course, in a real-world scenario, it would be necessary to determine the threshold using the training and validation data and then later apply that threshold to the similarity scores produced on unseen (test) data.

\rev{Our method of scoring the ``goodness'' of a model using $\text{mean(EER)} + 1.96 \times \text{SD(EER)}$ may have erroneously favored worse-performing models.}
\rev{It may be suggested for future works to exclude the SD term.}

\section{Conclusion}
We presented a metric learning approach for end-to-end biometric authentication via eye movements.
Our proposed model employed exponentially-dilated convolutions to exponentially increase the receptive field of subsequent layers while only linearly increasing the number of parameters.
Our approach was validated on the publicly available GazeBase dataset~\cite{Griffith2020} using recordings collected as much as 37~months apart, and we compared our approach against a statistical baseline and the current state-of-the-art, DEL.

When authenticating on R1 with a test-retest interval of approx. 20~minutes and using the first 10.240~s of each recording, we achieved an EER as low as 11.25\% on HSS @ 1000~Hz and a FRR @ FAR $10^{-4}$ as low as 55.77\% on TEX @ 50~Hz.
Even on R9 with a 37-month test-retest interval, we were able to achieve an EER as low as 18.15\% using the first 5.120~s of each recording (on a reduced pool of 14~subjects), but FRR @ FAR $10^{-4}$ was consistently above 99\%.

\rev{We have defined a testing scenario that is more realistic than most prior works, and we outperform the current state-of-the-art, DEL, under that testing scenario despite having 440x fewer learnable parameters and requiring a fraction of the time to train.}
\rev{Our work rehighlights the applicability of CNN architectures for eye movement biometrics and shows that even efficient architectures can achieve good performance.}
We believe that the scope and diversity (not only in terms of tasks, but also participant characteristics) of the GazeBase dataset gives it potential to serve as a unifying dataset for future biometrics research.
We also encourage future eye movement biometrics studies to report FRR @ FAR $10^{-4}$ in a combined effort to eventually achieve the FIDO Alliance's recommendation of 5\% FRR @ FAR $10^{-4}$.

\section*{Acknowledgment}
The authors would like to thank Dr. Lee Friedman for his suggestion of using the Pearson family of distributions for resampling similarities.
This material is based upon work supported by the National Science Foundation Graduate Research Fellowship under Grant No. DGE-1144466.
The study was also funded by 3 grants to Dr. Komogortsev: (1) National Science Foundation, CNS-1250718 and CNS-1714623, www.NSF.gov; (2) National Institute of Standards and Technology, 60NANB15D325, www.NIST.gov; (3) National Institute of Standards and Technology, 60NANB16D293.
Any opinions, findings, and conclusions or recommendations expressed in this material are those of the author(s) and do not necessarily reflect the views of the National Science Foundation or the National Institute of Standards and Technology.

\bibliographystyle{unsrt}
\bibliography{01main}

\section{Appendix}
\subsection{Hyperparameter Configurations}
During hyperparameter tuning and when training the final models, the workstation with RTX 2080 Tis processed the following configurations: TEX @ 1000~Hz, TEX @ 500~Hz, TEX @ 250~Hz, and TEX @ 125~Hz.
The workstation with RTX 3080s processed the following configurations: TEX @ 50~Hz, TEX @ 31.25~Hz, HSS, VD1, FXS, RAN, \rev{and BLG}.

See Tables~\ref{tab:search-space} and~\ref{tab:hparams}.

\begin{table}[htbp]
    \centering
    \caption{%
        Our search space for hyperparameter tuning. \\
        *: MS loss hyperparameters were not included while tuning the DEL baseline.%
    }
    \label{tab:search-space}
    \begin{tabular}{ll}
        \toprule
        Hyperparameter & Search bounds \\
        \midrule
        learning rate (log) & $[-6, -2]$ \\
        weight decay (log) & $[-6, -2]$ \\
        $\alpha$* & $[1, 100]$ \\
        $\beta$* & $[1, 100]$ \\
        $\lambda$* & $[0, 1]$ \\
        $\varepsilon$* & $[0, 0.5]$ \\
        \bottomrule
    \end{tabular}
\end{table}

\begin{table}[htbp]
    \centering
    \caption{%
        The set of hyperparameters used for each model, rounded to 2~decimal places.
        LR is learning rate and WD is weight decay. \\
        *: This set of hyperparameters was included for all models (except DEL) as part of the random (with a fixed seed) search phase.
        It happened to perform best for a majority of the models.%
    }
    \label{tab:hparams}
    \begin{tabular}{@{}lrrrrrr@{}}
        \toprule
        Model & log LR & log WD & $\alpha$ & $\beta$ & $\lambda$ & $\varepsilon$ \\
        \midrule
        HSS* & -2.12 & -3.17 & 6.75 & 60.51 & 0.87 & 0.01 \\
        VD1* & -2.12 & -3.17 & 6.75 & 60.51 & 0.87 & 0.01 \\
        FXS & -2.22 & -2.12 & 5.76 & 59.01 & 0.71 & 0.11 \\
        RAN* & -2.12 & -3.17 & 6.75 & 60.51 & 0.87 & 0.01 \\
        BLG* & -2.12 & -3.17 & 6.75 & 60.51 & 0.87 & 0.01 \\
        TEX @ 1000~Hz & -4.03 & -2.32 & 2.34 & 48.98 & 0.18 & 0.24 \\
        TEX @ 500~Hz & -2.29 & -5.61 & 13.28 & 57.26 & 0.19 & 0.40 \\
        TEX @ 250~Hz* & -2.12 & -3.17 & 6.75 & 60.51 & 0.87 & 0.01 \\
        TEX @ 125~Hz* & -2.12 & -3.17 & 6.75 & 60.51 & 0.87 & 0.01 \\
        TEX @ 50~Hz & -2.00 & -6.00 & 6.72 & 76.85 & 1.00 & 0.23 \\
        TEX @ 31.25~Hz & -2.00 & -2.00 & 13.01 & 35.66 & 1.00 & 0.50 \\
        DEL & \rev{-2.30} & \rev{-4.79} & -- & -- & -- & -- \\
        \bottomrule
    \end{tabular}
\end{table}

\subsection{Subjects per Fold}
See Table~\ref{tab:fold-subjects}.
\begin{table}[htbp]
    \centering
    \caption{The number of subjects present in each round for each split of the dataset.
    Subjects in subsequent rounds are a subset of the subjects present in the preceding round. \\
    *: subject~76 is missing from round~3. \\
    **: the total number of recordings present in the split which, since each subject was recorded twice within each round, is twice the sum of the row.}
    \label{tab:fold-subjects}
    \begin{tabular}{@{}lrrrrrrrrrr@{}}
        \toprule
        Split & R1 & R2 & R3 & R4 & R5 & R6 & R7 & R8 & R9 & N** \\
        \midrule
        F1 & 66 & 19 & 12  & 10 &  5 & -- & -- & -- & -- & 224 \\
        F2 & 66 & 19 & 11* & 10 &  6 & -- & -- & -- & -- & 224 \\
        F3 & 66 & 19 & 12  & 11 &  4 & -- & -- & -- & -- & 224 \\
        F4 & 65 & 20 & 11  & 11 &  4 & -- & -- & -- & -- & 222 \\
        Test  & 59 & 59 & 59  & 59 & 59 & 59 & 35 & 31 & 14 & 868 \\
        \midrule
        Total & 322 & 136 & 105 & 101 & 78 & 59 & 35 & 31 & 14 & 1762\\
        \bottomrule
    \end{tabular}
\end{table}

\subsection{Comprehensive Results for All Models}
These results are already presented in the main manuscript for TEX @ 1000~Hz, so we only present results for the other models here.
See Tables~\ref{tab:hss1000}--\ref{tab:blg1000} for tasks other than TEX.
See Tables~\ref{tab:tex500}--\ref{tab:tex31} for TEX at sampling rates other than 1000~Hz.
\rev{See Tables~\ref{tab:del}--\ref{tab:del-fast} for the DEL baseline including the slow and fast subnets individually.}
\rev{See Table~\ref{tab:star} for the STAR baseline.}

\begin{table*}
    \centering
    \caption{%
        Results on the held-out test set for HSS, varying $n$ and the authentication round.
        Values are presented as mean (standard deviation) across the 4~models trained with 4-fold cross-validation.%
    }
    \label{tab:hss1000}
    \begin{tabular}{lcccccc}
        \toprule
        \multirow{2}[2]{*}{$n$} & \multirow{2}[2]{*}{Round} & \multirow{2}[2]{*}{EER} & \multicolumn{4}{c}{FRR @ FAR} \\\cmidrule{4-7}
        {} & {} & {} & $10^{-1}$ & $10^{-2}$ & $10^{-3}$ & $10^{-4}$ \\
        \midrule
        \multirow{9}{*}{1} & 1 & 0.2407 (0.0142) & 0.4300 (0.0272) & 0.8347 (0.0253) & 0.9983 (0.0029) & 1.0000 (0.0000) \\
        &      2 & 0.3733 (0.0108) & 0.6913 (0.0330) & 0.9497 (0.0269) & 0.9999 (0.0001) & 1.0000 (0.0000) \\
        &      3 & 0.3694 (0.0227) & 0.7166 (0.0217) & 0.9530 (0.0192) & 0.9956 (0.0037) & 0.9998 (0.0004) \\
        &      4 & 0.3382 (0.0149) & 0.6867 (0.0279) & 0.9877 (0.0140) & 0.9998 (0.0003) & 1.0000 (0.0000) \\
        &      5 & 0.3485 (0.0080) & 0.7229 (0.0154) & 0.9823 (0.0148) & 0.9989 (0.0015) & 0.9999 (0.0002) \\
        &      6 & 0.3882 (0.0153) & 0.7273 (0.0277) & 0.9408 (0.0153) & 0.9954 (0.0054) & 0.9985 (0.0025) \\
        &      7 & 0.4072 (0.0217) & 0.8312 (0.0423) & 0.9874 (0.0172) & 0.9982 (0.0027) & 0.9997 (0.0005) \\
        &      8 & 0.3407 (0.0269) & 0.6910 (0.0325) & 0.9657 (0.0214) & 0.9998 (0.0003) & 1.0000 (0.0000) \\
        &      9 & 0.2802 (0.0186) & 0.5405 (0.0297) & 0.9556 (0.0340) & 0.9980 (0.0035) & 0.9999 (0.0002) \\\cmidrule{2-7}
        \multirow{9}{*}{5} & 1 & 0.1291 (0.0086) & 0.1640 (0.0173) & 0.5502 (0.0550) & 0.8431 (0.0784) & 0.9589 (0.0413) \\
        &      2 & 0.2226 (0.0034) & 0.3995 (0.0251) & 0.7671 (0.0277) & 0.9346 (0.0139) & 0.9847 (0.0072) \\
        &      3 & 0.2381 (0.0029) & 0.4563 (0.0147) & 0.8150 (0.0264) & 0.9507 (0.0116) & 0.9859 (0.0036) \\
        &      4 & 0.2127 (0.0126) & 0.3793 (0.0313) & 0.7757 (0.0148) & 0.9418 (0.0121) & 0.9862 (0.0119) \\
        &      5 & 0.1961 (0.0040) & 0.3352 (0.0123) & 0.7620 (0.0249) & 0.9521 (0.0148) & 0.9958 (0.0052) \\
        &      6 & 0.2383 (0.0076) & 0.4727 (0.0163) & 0.8476 (0.0246) & 0.9589 (0.0153) & 0.9845 (0.0088) \\
        &      7 & 0.2529 (0.0028) & 0.4930 (0.0276) & 0.8391 (0.0386) & 0.9632 (0.0276) & 0.9890 (0.0152) \\
        &      8 & 0.2323 (0.0125) & 0.4528 (0.0144) & 0.8792 (0.0280) & 1.0000 (0.0000) & 1.0000 (0.0000) \\
        &      9 & 0.1638 (0.0052) & 0.2503 (0.0284) & 0.6804 (0.0776) & 0.9547 (0.0785) & 0.9667 (0.0576) \\\cmidrule{2-7}
        \multirow{9}{*}{10} & 1 & 0.1125 (0.0071) & 0.1273 (0.0154) & 0.4867 (0.0299) & 0.8010 (0.0891) & 0.9343 (0.0542) \\
       &      2 & 0.2112 (0.0031) & 0.3712 (0.0111) & 0.7294 (0.0123) & 0.9004 (0.0283) & 0.9556 (0.0279) \\
       &      3 & 0.2283 (0.0039) & 0.4376 (0.0106) & 0.7885 (0.0190) & 0.9152 (0.0156) & 0.9610 (0.0191) \\
       &      4 & 0.2091 (0.0104) & 0.3642 (0.0246) & 0.7351 (0.0204) & 0.9072 (0.0093) & 0.9709 (0.0245) \\
       &      5 & 0.1804 (0.0043) & 0.2990 (0.0118) & 0.7178 (0.0382) & 0.9149 (0.0444) & 0.9783 (0.0188) \\
       &      6 & 0.2310 (0.0082) & 0.4630 (0.0174) & 0.8377 (0.0302) & 0.9495 (0.0210) & 0.9808 (0.0096) \\
       &      7 & 0.2476 (0.0077) & 0.4732 (0.0213) & 0.8044 (0.0332) & 0.9258 (0.0239) & 0.9633 (0.0231) \\
       &      8 & 0.2233 (0.0087) & 0.4237 (0.0111) & 0.8216 (0.0236) & 0.9944 (0.0056) & 1.0000 (0.0000) \\
       &      9 & 0.1499 (0.0121) & 0.2068 (0.0259) & 0.5750 (0.0669) & 0.9585 (0.0452) & 0.9990 (0.0018) \\
        \bottomrule
    \end{tabular}
\end{table*}

\begin{table*}
    \centering
    \caption{%
        Results on the held-out test set for VD1, varying $n$ and the authentication round.
        Values are presented as mean (standard deviation) across the 4~models trained with 4-fold cross-validation.%
    }
    \label{tab:vd11000}
    \begin{tabular}{lcccccc}
        \toprule
        \multirow{2}[2]{*}{$n$} & \multirow{2}[2]{*}{Round} & \multirow{2}[2]{*}{EER} & \multicolumn{4}{c}{FRR @ FAR} \\\cmidrule{4-7}
        {} & {} & {} & $10^{-1}$ & $10^{-2}$ & $10^{-3}$ & $10^{-4}$ \\
        \midrule
        \multirow{9}{*}{1} & 1 & 0.2714 (0.0150) & 0.5072 (0.0405) & 0.8722 (0.0169) & 0.9920 (0.0049) & 0.9994 (0.0010) \\
        &      2 & 0.3565 (0.0211) & 0.6905 (0.0292) & 0.9497 (0.0127) & 0.9957 (0.0035) & 0.9997 (0.0002) \\
        &      3 & 0.3562 (0.0038) & 0.7247 (0.0208) & 0.9536 (0.0171) & 0.9941 (0.0066) & 0.9981 (0.0025) \\
        &      4 & 0.3622 (0.0128) & 0.7573 (0.0253) & 0.9559 (0.0131) & 0.9882 (0.0059) & 0.9948 (0.0031) \\
        &      5 & 0.3249 (0.0119) & 0.6288 (0.0293) & 0.9376 (0.0199) & 0.9961 (0.0046) & 0.9992 (0.0011) \\
        &      6 & 0.3798 (0.0107) & 0.7496 (0.0150) & 0.9442 (0.0087) & 0.9833 (0.0047) & 0.9930 (0.0032) \\
        &      7 & 0.3369 (0.0355) & 0.6413 (0.0379) & 0.9118 (0.0152) & 0.9857 (0.0079) & 0.9984 (0.0010) \\
        &      8 & 0.3713 (0.0270) & 0.7165 (0.0384) & 0.9190 (0.0160) & 0.9843 (0.0062) & 0.9988 (0.0013) \\
        &      9 & 0.3057 (0.0241) & 0.5431 (0.0413) & 0.8051 (0.0464) & 0.9146 (0.0656) & 0.9471 (0.0704) \\\cmidrule{2-7}
        \multirow{9}{*}{5} & 1 & 0.1888 (0.0110) & 0.3210 (0.0284) & 0.7470 (0.0450) & 0.9344 (0.0453) & 0.9822 (0.0259) \\
        &      2 & 0.2912 (0.0126) & 0.5730 (0.0382) & 0.8851 (0.0179) & 0.9709 (0.0172) & 0.9895 (0.0106) \\
        &      3 & 0.3253 (0.0147) & 0.6375 (0.0460) & 0.9267 (0.0344) & 0.9836 (0.0125) & 0.9961 (0.0030) \\
        &      4 & 0.3000 (0.0155) & 0.5800 (0.0352) & 0.8758 (0.0088) & 0.9661 (0.0124) & 0.9879 (0.0071) \\
        &      5 & 0.2852 (0.0102) & 0.5632 (0.0366) & 0.8867 (0.0308) & 0.9704 (0.0153) & 0.9870 (0.0085) \\
        &      6 & 0.3175 (0.0131) & 0.6173 (0.0372) & 0.9253 (0.0077) & 0.9957 (0.0073) & 0.9989 (0.0020) \\
        &      7 & 0.2699 (0.0196) & 0.5160 (0.0388) & 0.8904 (0.0122) & 0.9900 (0.0085) & 0.9988 (0.0015) \\
        &      8 & 0.3116 (0.0057) & 0.6023 (0.0061) & 0.8537 (0.0221) & 0.9423 (0.0163) & 0.9727 (0.0179) \\
        &      9 & 0.2064 (0.0193) & 0.4012 (0.0709) & 0.8262 (0.0744) & 0.9680 (0.0342) & 0.9939 (0.0087) \\\cmidrule{2-7}
        \multirow{9}{*}{10} & 1 & 0.1694 (0.0112) & 0.2770 (0.0276) & 0.7343 (0.0460) & 0.9288 (0.0498) & 0.9789 (0.0298) \\
       &      2 & 0.2876 (0.0184) & 0.5565 (0.0281) & 0.8766 (0.0229) & 0.9677 (0.0201) & 0.9891 (0.0122) \\
       &      3 & 0.3279 (0.0107) & 0.6283 (0.0128) & 0.8858 (0.0277) & 0.9630 (0.0242) & 0.9850 (0.0124) \\
       &      4 & 0.2849 (0.0146) & 0.5667 (0.0304) & 0.8642 (0.0063) & 0.9551 (0.0141) & 0.9755 (0.0100) \\
       &      5 & 0.2746 (0.0174) & 0.5492 (0.0279) & 0.8937 (0.0220) & 0.9696 (0.0161) & 0.9879 (0.0100) \\
       &      6 & 0.3151 (0.0179) & 0.6099 (0.0297) & 0.9125 (0.0016) & 0.9901 (0.0080) & 0.9979 (0.0030) \\
       &      7 & 0.2741 (0.0222) & 0.5343 (0.0387) & 0.8630 (0.0367) & 0.9605 (0.0200) & 0.9864 (0.0105) \\
       &      8 & 0.2769 (0.0117) & 0.5638 (0.0286) & 0.8586 (0.0340) & 0.9472 (0.0258) & 0.9709 (0.0182) \\
       &      9 & 0.1971 (0.0138) & 0.3937 (0.0447) & 0.8467 (0.0660) & 0.9620 (0.0426) & 0.9847 (0.0237) \\
        \bottomrule
    \end{tabular}
\end{table*}

\begin{table*}
    \centering
    \caption{%
        Results on the held-out test set for FXS, varying $n$ and the authentication round.
        Values are presented as mean (standard deviation) across the 4~models trained with 4-fold cross-validation.%
    }
    \label{tab:fxs1000}
    \begin{tabular}{lcccccc}
        \toprule
        \multirow{2}[2]{*}{$n$} & \multirow{2}[2]{*}{Round} & \multirow{2}[2]{*}{EER} & \multicolumn{4}{c}{FRR @ FAR} \\\cmidrule{4-7}
        {} & {} & {} & $10^{-1}$ & $10^{-2}$ & $10^{-3}$ & $10^{-4}$ \\
        \midrule
        \multirow{9}{*}{1} & 1 & 0.2740 (0.0109) & 0.5252 (0.0160) & 0.8934 (0.0098) & 0.9890 (0.0097) & 0.9975 (0.0034) \\
        &      2 & 0.3722 (0.0266) & 0.7469 (0.0423) & 0.9858 (0.0081) & 0.9996 (0.0006) & 1.0000 (0.0001) \\
        &      3 & 0.3943 (0.0194) & 0.7362 (0.0125) & 0.9510 (0.0050) & 0.9919 (0.0064) & 0.9976 (0.0034) \\
        &      4 & 0.3761 (0.0150) & 0.7203 (0.0229) & 0.9457 (0.0158) & 0.9919 (0.0074) & 0.9983 (0.0025) \\
        &      5 & 0.3526 (0.0091) & 0.6911 (0.0075) & 0.9449 (0.0036) & 0.9929 (0.0027) & 0.9991 (0.0005) \\
        &      6 & 0.3876 (0.0123) & 0.7422 (0.0260) & 0.9606 (0.0092) & 0.9968 (0.0034) & 0.9997 (0.0003) \\
        &      7 & 0.3786 (0.0142) & 0.7245 (0.0188) & 0.9470 (0.0011) & 0.9922 (0.0056) & 0.9977 (0.0033) \\
        &      8 & 0.3652 (0.0241) & 0.7347 (0.0373) & 0.9658 (0.0062) & 0.9973 (0.0017) & 0.9997 (0.0002) \\
        &      9 & 0.3797 (0.0318) & 0.7296 (0.0707) & 0.9278 (0.0232) & 0.9842 (0.0071) & 0.9977 (0.0016) \\\cmidrule{2-7}
        \multirow{9}{*}{5} & 1 & 0.2198 (0.0175) & 0.4493 (0.0470) & 0.8794 (0.0215) & 0.9812 (0.0149) & 0.9962 (0.0048) \\
        &      2 & 0.3365 (0.0035) & 0.6763 (0.0217) & 0.9533 (0.0041) & 0.9975 (0.0022) & 0.9998 (0.0001) \\
        &      3 & 0.3510 (0.0088) & 0.6831 (0.0058) & 0.9495 (0.0031) & 0.9971 (0.0038) & 0.9995 (0.0009) \\
        &      4 & 0.3577 (0.0209) & 0.7157 (0.0215) & 0.9397 (0.0165) & 0.9859 (0.0082) & 0.9956 (0.0031) \\
        &      5 & 0.3146 (0.0089) & 0.6305 (0.0164) & 0.9272 (0.0151) & 0.9903 (0.0041) & 0.9988 (0.0005) \\
        &      6 & 0.3741 (0.0103) & 0.7301 (0.0161) & 0.9594 (0.0076) & 0.9967 (0.0026) & 0.9997 (0.0002) \\
        &      7 & 0.3502 (0.0196) & 0.7337 (0.0288) & 0.9518 (0.0147) & 0.9887 (0.0065) & 0.9969 (0.0029) \\
        &      8 & 0.3347 (0.0166) & 0.6892 (0.0283) & 0.9487 (0.0126) & 0.9958 (0.0025) & 0.9995 (0.0005) \\
        &      9 & 0.3143 (0.0091) & 0.6625 (0.0182) & 0.9376 (0.0190) & 0.9902 (0.0070) & 0.9959 (0.0061) \\\cmidrule{2-7}
        \multirow{9}{*}{10} & 1 & 0.2136 (0.0102) & 0.4260 (0.0389) & 0.8664 (0.0329) & 0.9763 (0.0205) & 0.9953 (0.0064) \\
       &      2 & 0.3367 (0.0059) & 0.6796 (0.0229) & 0.9577 (0.0074) & 0.9965 (0.0030) & 0.9991 (0.0011) \\
       &      3 & 0.3545 (0.0126) & 0.6805 (0.0147) & 0.9554 (0.0210) & 0.9953 (0.0057) & 0.9993 (0.0010) \\
       &      4 & 0.3526 (0.0215) & 0.7100 (0.0188) & 0.9356 (0.0165) & 0.9845 (0.0082) & 0.9952 (0.0029) \\
       &      5 & 0.3111 (0.0119) & 0.6289 (0.0056) & 0.9360 (0.0129) & 0.9915 (0.0045) & 0.9991 (0.0008) \\
       &      6 & 0.3654 (0.0107) & 0.7113 (0.0218) & 0.9518 (0.0040) & 0.9949 (0.0021) & 0.9994 (0.0005) \\
       &      7 & 0.3299 (0.0202) & 0.7112 (0.0310) & 0.9571 (0.0119) & 0.9922 (0.0031) & 0.9988 (0.0013) \\
       &      8 & 0.3344 (0.0160) & 0.6726 (0.0265) & 0.9501 (0.0075) & 0.9960 (0.0033) & 0.9996 (0.0004) \\
       &      9 & 0.2931 (0.0167) & 0.6106 (0.0366) & 0.9470 (0.0336) & 0.9908 (0.0138) & 0.9951 (0.0082) \\
        \bottomrule
    \end{tabular}
\end{table*}

\begin{table*}
    \centering
    \caption{%
        Results on the held-out test set for RAN, varying $n$ and the authentication round.
        Values are presented as mean (standard deviation) across the 4~models trained with 4-fold cross-validation.%
    }
    \label{tab:ran1000}
    \begin{tabular}{lcccccc}
        \toprule
        \multirow{2}[2]{*}{$n$} & \multirow{2}[2]{*}{Round} & \multirow{2}[2]{*}{EER} & \multicolumn{4}{c}{FRR @ FAR} \\\cmidrule{4-7}
        {} & {} & {} & $10^{-1}$ & $10^{-2}$ & $10^{-3}$ & $10^{-4}$ \\
        \midrule
        \multirow{9}{*}{1} & 1 & 0.2091 (0.0090) & 0.3809 (0.0316) & 0.8292 (0.0337) & 0.9912 (0.0070) & 0.9997 (0.0003) \\
        &      2 & 0.2806 (0.0204) & 0.5517 (0.0293) & 0.9246 (0.0322) & 0.9996 (0.0008) & 1.0000 (0.0000) \\
        &      3 & 0.2981 (0.0241) & 0.5857 (0.0638) & 0.9189 (0.0284) & 0.9963 (0.0053) & 1.0000 (0.0000) \\
        &      4 & 0.2891 (0.0193) & 0.5671 (0.0393) & 0.9200 (0.0413) & 0.9909 (0.0078) & 0.9989 (0.0018) \\
        &      5 & 0.2927 (0.0201) & 0.5793 (0.0459) & 0.9161 (0.0098) & 0.9906 (0.0100) & 0.9982 (0.0032) \\
        &      6 & 0.3210 (0.0091) & 0.6510 (0.0268) & 0.9723 (0.0182) & 0.9988 (0.0021) & 1.0000 (0.0000) \\
        &      7 & 0.3160 (0.0077) & 0.6221 (0.0271) & 0.9184 (0.0157) & 0.9935 (0.0065) & 0.9993 (0.0008) \\
        &      8 & 0.2734 (0.0114) & 0.5403 (0.0382) & 0.9327 (0.0385) & 0.9918 (0.0111) & 0.9997 (0.0005) \\
        &      9 & 0.2883 (0.0216) & 0.5841 (0.0613) & 0.9662 (0.0446) & 0.9995 (0.0009) & 0.9999 (0.0001) \\\cmidrule{2-7}
        \multirow{9}{*}{5} & 1 & 0.1450 (0.0047) & 0.2208 (0.0125) & 0.7568 (0.0649) & 0.9591 (0.0393) & 0.9918 (0.0111) \\
        &      2 & 0.2410 (0.0238) & 0.4824 (0.0395) & 0.9023 (0.0204) & 0.9948 (0.0054) & 1.0000 (0.0000) \\
        &      3 & 0.2717 (0.0215) & 0.5380 (0.0496) & 0.8908 (0.0156) & 0.9822 (0.0119) & 0.9945 (0.0046) \\
        &      4 & 0.2471 (0.0221) & 0.5014 (0.0594) & 0.8715 (0.0444) & 0.9673 (0.0214) & 0.9878 (0.0124) \\
        &      5 & 0.2411 (0.0114) & 0.4932 (0.0399) & 0.8709 (0.0327) & 0.9709 (0.0138) & 0.9903 (0.0074) \\
        &      6 & 0.2669 (0.0215) & 0.5377 (0.0410) & 0.9339 (0.0121) & 0.9966 (0.0058) & 0.9993 (0.0012) \\
        &      7 & 0.2634 (0.0129) & 0.5432 (0.0372) & 0.8933 (0.0343) & 0.9783 (0.0160) & 0.9936 (0.0082) \\
        &      8 & 0.2478 (0.0090) & 0.4833 (0.0206) & 0.9178 (0.0201) & 0.9989 (0.0018) & 1.0000 (0.0000) \\
        &      9 & 0.1988 (0.0252) & 0.3659 (0.0077) & 0.9545 (0.0339) & 0.9993 (0.0012) & 1.0000 (0.0000) \\\cmidrule{2-7}
        \multirow{9}{*}{10} & 1 & 0.1459 (0.0079) & 0.2293 (0.0210) & 0.7602 (0.0619) & 0.9481 (0.0445) & 0.9836 (0.0175) \\
       &      2 & 0.2376 (0.0250) & 0.4762 (0.0368) & 0.8883 (0.0167) & 0.9914 (0.0069) & 0.9978 (0.0037) \\
       &      3 & 0.2697 (0.0217) & 0.5410 (0.0427) & 0.8678 (0.0226) & 0.9644 (0.0183) & 0.9873 (0.0114) \\
       &      4 & 0.2408 (0.0196) & 0.4750 (0.0492) & 0.8505 (0.0399) & 0.9578 (0.0235) & 0.9805 (0.0201) \\
       &      5 & 0.2348 (0.0168) & 0.4850 (0.0507) & 0.8700 (0.0399) & 0.9690 (0.0214) & 0.9879 (0.0104) \\
       &      6 & 0.2626 (0.0194) & 0.5294 (0.0385) & 0.9188 (0.0160) & 0.9927 (0.0081) & 0.9987 (0.0017) \\
       &      7 & 0.2552 (0.0146) & 0.5158 (0.0322) & 0.8867 (0.0292) & 0.9778 (0.0239) & 0.9934 (0.0088) \\
       &      8 & 0.2413 (0.0077) & 0.4784 (0.0175) & 0.9167 (0.0288) & 0.9992 (0.0008) & 1.0000 (0.0000) \\
       &      9 & 0.1888 (0.0190) & 0.3639 (0.0420) & 0.9288 (0.0511) & 0.9976 (0.0040) & 0.9997 (0.0005) \\
        \bottomrule
    \end{tabular}
\end{table*}

\begin{table*}
    \centering
    \caption{%
        Results on the held-out test set for BLG, varying $n$ and the authentication round.
        Values are presented as mean (standard deviation) across the 4~models trained with 4-fold cross-validation.%
    }
    \label{tab:blg1000}
    \begin{tabular}{lcccccc}
        \toprule
        \multirow{2}[2]{*}{$n$} & \multirow{2}[2]{*}{Round} & \multirow{2}[2]{*}{EER} & \multicolumn{4}{c}{FRR @ FAR} \\\cmidrule{4-7}
        {} & {} & {} & $10^{-1}$ & $10^{-2}$ & $10^{-3}$ & $10^{-4}$ \\
        \midrule
        \multirow{9}{*}{1} & 1 & 0.2328 (0.0189) & 0.4328 (0.0184) & 0.8242 (0.0422) & 0.9829 (0.0208) & 0.9980 (0.0030) \\
        &      2 & 0.3034 (0.0106) & 0.5641 (0.0231) & 0.8749 (0.0307) & 0.9972 (0.0032) & 1.0000 (0.0000) \\
        &      3 & 0.3463 (0.0232) & 0.6845 (0.0513) & 0.9622 (0.0306) & 0.9992 (0.0009) & 1.0000 (0.0000) \\
        &      4 & 0.3132 (0.0172) & 0.6287 (0.0302) & 0.9425 (0.0187) & 0.9978 (0.0023) & 1.0000 (0.0000) \\
        &      5 & 0.3145 (0.0183) & 0.6163 (0.0193) & 0.9430 (0.0213) & 0.9970 (0.0031) & 0.9999 (0.0001) \\
        &      6 & 0.3606 (0.0125) & 0.7026 (0.0283) & 0.9588 (0.0108) & 0.9982 (0.0018) & 0.9999 (0.0001) \\
        &      7 & 0.3543 (0.0229) & 0.6729 (0.0526) & 0.9563 (0.0185) & 0.9968 (0.0055) & 0.9990 (0.0018) \\
        &      8 & 0.3235 (0.0178) & 0.5732 (0.0187) & 0.8760 (0.0440) & 0.9994 (0.0010) & 1.0000 (0.0000) \\
        &      9 & 0.3362 (0.0648) & 0.6321 (0.0536) & 0.9475 (0.0410) & 0.9953 (0.0078) & 0.9983 (0.0029) \\\cmidrule{2-7}
        \multirow{9}{*}{5} & 1 & 0.1517 (0.0138) & 0.2478 (0.0309) & 0.7191 (0.0191) & 0.9246 (0.0261) & 0.9810 (0.0164) \\
        &      2 & 0.2500 (0.0136) & 0.4522 (0.0195) & 0.8640 (0.0197) & 1.0000 (0.0000) & 1.0000 (0.0000) \\
        &      3 & 0.2835 (0.0148) & 0.5474 (0.0384) & 0.8758 (0.0105) & 0.9793 (0.0144) & 0.9959 (0.0035) \\
        &      4 & 0.2511 (0.0116) & 0.5242 (0.0321) & 0.8560 (0.0116) & 0.9561 (0.0087) & 0.9787 (0.0132) \\
        &      5 & 0.2540 (0.0226) & 0.4865 (0.0398) & 0.8405 (0.0291) & 0.9488 (0.0333) & 0.9718 (0.0220) \\
        &      6 & 0.2916 (0.0134) & 0.5914 (0.0239) & 0.8867 (0.0086) & 0.9639 (0.0137) & 0.9816 (0.0083) \\
        &      7 & 0.2782 (0.0132) & 0.5668 (0.0198) & 0.8966 (0.0197) & 0.9768 (0.0151) & 0.9897 (0.0075) \\
        &      8 & 0.2373 (0.0116) & 0.4715 (0.0267) & 0.8836 (0.0151) & 0.9944 (0.0071) & 1.0000 (0.0001) \\
        &      9 & 0.2001 (0.0133) & 0.3480 (0.0383) & 0.8198 (0.0732) & 0.9835 (0.0209) & 0.9998 (0.0004) \\\cmidrule{2-7}
        \multirow{9}{*}{10} & 1 & 0.1404 (0.0112) & 0.2140 (0.0273) & 0.6835 (0.0125) & 0.8902 (0.0355) & 0.9488 (0.0272) \\
       &      2 & 0.2381 (0.0134) & 0.4389 (0.0277) & 0.8301 (0.0222) & 0.9715 (0.0282) & 0.9949 (0.0052) \\
       &      3 & 0.2751 (0.0123) & 0.5204 (0.0303) & 0.8324 (0.0141) & 0.9523 (0.0171) & 0.9875 (0.0141) \\
       &      4 & 0.2380 (0.0148) & 0.4660 (0.0482) & 0.8104 (0.0163) & 0.9337 (0.0170) & 0.9682 (0.0190) \\
       &      5 & 0.2344 (0.0228) & 0.4339 (0.0526) & 0.7953 (0.0248) & 0.9218 (0.0324) & 0.9542 (0.0271) \\
       &      6 & 0.2833 (0.0132) & 0.5578 (0.0313) & 0.8490 (0.0181) & 0.9315 (0.0282) & 0.9528 (0.0330) \\
       &      7 & 0.2717 (0.0138) & 0.5551 (0.0128) & 0.8774 (0.0233) & 0.9534 (0.0216) & 0.9787 (0.0141) \\
       &      8 & 0.2194 (0.0106) & 0.4137 (0.0190) & 0.8395 (0.0205) & 0.9920 (0.0087) & 1.0000 (0.0000) \\
       &      9 & 0.1784 (0.0148) & 0.2885 (0.0339) & 0.7339 (0.0491) & 0.9588 (0.0417) & 0.9921 (0.0118) \\
        \bottomrule
    \end{tabular}
\end{table*}

\begin{table*}
    \centering
    \caption{%
        Results on the held-out test set for TEX @ 500~Hz, varying $n$ and the authentication round.
        Values are presented as mean (standard deviation) across the 4~models trained with 4-fold cross-validation.%
    }
    \label{tab:tex500}
    \begin{tabular}{lcccccc}
        \toprule
        \multirow{2}[2]{*}{$n$} & \multirow{2}[2]{*}{Round} & \multirow{2}[2]{*}{EER} & \multicolumn{4}{c}{FRR @ FAR} \\\cmidrule{4-7}
        {} & {} & {} & $10^{-1}$ & $10^{-2}$ & $10^{-3}$ & $10^{-4}$ \\
        \midrule
        \multirow{9}{*}{1} & 1 & 0.2672 (0.0156) & 0.4898 (0.0557) & 0.8203 (0.0503) & 0.9593 (0.0221) & 0.9972 (0.0027) \\
        {} & 2 & 0.3492 (0.0041) & 0.6811 (0.0434) & 0.9414 (0.0205) & 0.9864 (0.0099) & 0.9967 (0.0028) \\
        {} & 3 & 0.3487 (0.0119) & 0.6960 (0.0422) & 0.9262 (0.0264) & 0.9803 (0.0115) & 0.9943 (0.0042) \\
        {} & 4 & 0.3749 (0.0179) & 0.7195 (0.0312) & 0.9325 (0.0169) & 0.9839 (0.0068) & 0.9950 (0.0022) \\
        {} & 5 & 0.3176 (0.0155) & 0.5794 (0.0283) & 0.8684 (0.0089) & 0.9701 (0.0072) & 0.9935 (0.0056) \\
        {} & 6 & 0.3402 (0.0153) & 0.6375 (0.0251) & 0.9009 (0.0064) & 0.9797 (0.0014) & 0.9959 (0.0015) \\
        {} & 7 & 0.3711 (0.0296) & 0.6725 (0.0553) & 0.8873 (0.0414) & 0.9594 (0.0195) & 0.9891 (0.0079) \\
        {} & 8 & 0.3347 (0.0371) & 0.6275 (0.0623) & 0.9398 (0.0350) & 0.9972 (0.0043) & 0.9998 (0.0004) \\
        {} & 9 & 0.3491 (0.0551) & 0.6711 (0.0723) & 0.9057 (0.0515) & 0.9780 (0.0139) & 0.9956 (0.0074) \\\cmidrule{2-7}
        \multirow{9}{*}{5} & 1 & 0.1768 (0.0144) & 0.3276 (0.0518) & 0.8253 (0.0323) & 0.9893 (0.0072) & 0.9996 (0.0005) \\
        {} & 2 & 0.2524 (0.0144) & 0.5025 (0.0395) & 0.9221 (0.0296) & 0.9944 (0.0079) & 0.9995 (0.0008) \\
        {} & 3 & 0.2506 (0.0264) & 0.4895 (0.0444) & 0.8959 (0.0239) & 0.9914 (0.0051) & 0.9997 (0.0003) \\
        {} & 4 & 0.2675 (0.0220) & 0.5185 (0.0490) & 0.8407 (0.0382) & 0.9462 (0.0186) & 0.9822 (0.0071) \\
        {} & 5 & 0.2503 (0.0160) & 0.4768 (0.0174) & 0.8781 (0.0363) & 0.9879 (0.0086) & 0.9996 (0.0003) \\
        {} & 6 & 0.2765 (0.0154) & 0.5496 (0.0246) & 0.9313 (0.0158) & 0.9984 (0.0013) & 0.9999 (0.0001) \\
        {} & 7 & 0.2730 (0.0110) & 0.5457 (0.0541) & 0.8816 (0.0374) & 0.9882 (0.0054) & 0.9996 (0.0004) \\
        {} & 8 & 0.2865 (0.0178) & 0.5802 (0.0167) & 0.9339 (0.0122) & 0.9996 (0.0004) & 1.0000 (0.0000) \\
        {} & 9 & 0.2302 (0.0411) & 0.4933 (0.1107) & 0.8605 (0.0361) & 0.9762 (0.0128) & 0.9973 (0.0033) \\\cmidrule{2-7}
        \multirow{9}{*}{10} & 1 & 0.1667 (0.0155) & 0.3110 (0.0577) & 0.8315 (0.0293) & 0.9928 (0.0051) & 1.0000 (0.0000) \\
        {} & 2 & 0.2435 (0.0184) & 0.4866 (0.0503) & 0.9192 (0.0341) & 0.9932 (0.0059) & 0.9997 (0.0004) \\
        {} & 3 & 0.2487 (0.0183) & 0.4803 (0.0366) & 0.8832 (0.0251) & 0.9893 (0.0061) & 0.9998 (0.0002) \\
        {} & 4 & 0.2552 (0.0191) & 0.4917 (0.0407) & 0.8163 (0.0282) & 0.9317 (0.0156) & 0.9754 (0.0132) \\
        {} & 5 & 0.2478 (0.0212) & 0.4878 (0.0510) & 0.8733 (0.0409) & 0.9898 (0.0063) & 0.9999 (0.0000) \\
        {} & 6 & 0.2764 (0.0158) & 0.5618 (0.0317) & 0.9280 (0.0179) & 0.9975 (0.0020) & 0.9999 (0.0001) \\
        {} & 7 & 0.2542 (0.0102) & 0.4989 (0.0465) & 0.8863 (0.0275) & 0.9940 (0.0031) & 1.0000 (0.0001) \\
        {} & 8 & 0.2635 (0.0229) & 0.5723 (0.0358) & 0.9558 (0.0125) & 0.9996 (0.0004) & 1.0000 (0.0000) \\
        {} & 9 & 0.2344 (0.0345) & 0.4608 (0.1046) & 0.8097 (0.0626) & 0.9569 (0.0184) & 0.9953 (0.0048) \\
        \bottomrule
    \end{tabular}
\end{table*}

\begin{table*}
    \centering
    \caption{%
        Results on the held-out test set for TEX @ 250~Hz, varying $n$ and the authentication round.
        Values are presented as mean (standard deviation) across the 4~models trained with 4-fold cross-validation.%
    }
    \label{tab:tex250}
    \begin{tabular}{lcccccc}
        \toprule
        \multirow{2}[2]{*}{$n$} & \multirow{2}[2]{*}{Round} & \multirow{2}[2]{*}{EER} & \multicolumn{4}{c}{FRR @ FAR} \\\cmidrule{4-7}
        {} & {} & {} & $10^{-1}$ & $10^{-2}$ & $10^{-3}$ & $10^{-4}$ \\
        \midrule
        \multirow{9}{*}{1} & 1 & 0.2697 (0.0058) & 0.5096 (0.0331) & 0.8825 (0.0469) & 0.9988 (0.0011) & 1.0000 (0.0001) \\
        {} &      2 & 0.3140 (0.0128) & 0.6119 (0.0285) & 0.9305 (0.0131) & 0.9986 (0.0024) & 1.0000 (0.0001) \\
        {} &      3 & 0.3201 (0.0122) & 0.6599 (0.0373) & 0.9556 (0.0274) & 0.9976 (0.0025) & 0.9996 (0.0007) \\
        {} &      4 & 0.3328 (0.0261) & 0.6678 (0.0387) & 0.9631 (0.0116) & 0.9999 (0.0001) & 1.0000 (0.0000) \\
        {} &      5 & 0.2927 (0.0118) & 0.5694 (0.0466) & 0.9032 (0.0348) & 0.9974 (0.0038) & 0.9998 (0.0004) \\
        {} &      6 & 0.3034 (0.0211) & 0.6059 (0.0307) & 0.9473 (0.0433) & 0.9933 (0.0117) & 0.9976 (0.0041) \\
        {} &      7 & 0.3220 (0.0076) & 0.6397 (0.0057) & 0.9304 (0.0224) & 0.9841 (0.0165) & 0.9916 (0.0106) \\
        {} &      8 & 0.2831 (0.0156) & 0.5568 (0.0474) & 0.8983 (0.0366) & 0.9856 (0.0142) & 0.9973 (0.0034) \\
        {} &      9 & 0.2913 (0.0180) & 0.5783 (0.0617) & 0.9529 (0.0273) & 1.0000 (0.0000) & 1.0000 (0.0000) \\\cmidrule{2-7}
        \multirow{9}{*}{5} & 1 & 0.1684 (0.0074) & 0.2662 (0.0209) & 0.7191 (0.0579) & 0.9525 (0.0504) & 0.9930 (0.0114) \\
         &      2 & 0.2313 (0.0168) & 0.4047 (0.0302) & 0.8296 (0.0472) & 0.9902 (0.0109) & 1.0000 (0.0000) \\
         &      3 & 0.2600 (0.0154) & 0.4471 (0.0292) & 0.8053 (0.0285) & 0.9806 (0.0188) & 0.9986 (0.0024) \\
         &      4 & 0.2578 (0.0223) & 0.4609 (0.0389) & 0.8089 (0.0199) & 0.9737 (0.0196) & 0.9971 (0.0031) \\
         &      5 & 0.2540 (0.0096) & 0.4758 (0.0232) & 0.8381 (0.0225) & 0.9783 (0.0144) & 0.9972 (0.0028) \\
         &      6 & 0.2735 (0.0195) & 0.4997 (0.0272) & 0.8687 (0.0305) & 0.9833 (0.0206) & 0.9977 (0.0039) \\
         &      7 & 0.2541 (0.0067) & 0.4541 (0.0329) & 0.8025 (0.0570) & 0.9679 (0.0419) & 0.9944 (0.0097) \\
         &      8 & 0.2607 (0.0158) & 0.4941 (0.0394) & 0.8515 (0.0259) & 0.9903 (0.0099) & 0.9996 (0.0006) \\
         &      9 & 0.2342 (0.0185) & 0.3911 (0.0277) & 0.8190 (0.0883) & 0.9924 (0.0131) & 1.0000 (0.0000) \\\cmidrule{2-7}
        \multirow{9}{*}{10} & 1 & 0.1661 (0.0091) & 0.2540 (0.0201) & 0.6916 (0.0439) & 0.9597 (0.0399) & 0.9999 (0.0001) \\
       &      2 & 0.2181 (0.0211) & 0.3876 (0.0462) & 0.8282 (0.0348) & 0.9908 (0.0139) & 0.9996 (0.0006) \\
       &      3 & 0.2528 (0.0124) & 0.4340 (0.0205) & 0.7885 (0.0371) & 0.9717 (0.0180) & 1.0000 (0.0000) \\
       &      4 & 0.2312 (0.0193) & 0.4174 (0.0395) & 0.7969 (0.0476) & 0.9594 (0.0307) & 0.9906 (0.0076) \\
       &      5 & 0.2281 (0.0083) & 0.4202 (0.0320) & 0.8153 (0.0398) & 0.9648 (0.0229) & 0.9936 (0.0063) \\
       &      6 & 0.2660 (0.0157) & 0.4845 (0.0280) & 0.8560 (0.0304) & 0.9790 (0.0196) & 0.9959 (0.0067) \\
       &      7 & 0.2475 (0.0061) & 0.4305 (0.0268) & 0.7834 (0.0471) & 0.9592 (0.0425) & 0.9923 (0.0086) \\
       &      8 & 0.2454 (0.0146) & 0.4833 (0.0395) & 0.8767 (0.0281) & 0.9933 (0.0043) & 1.0000 (0.0000) \\
       &      9 & 0.2287 (0.0144) & 0.3654 (0.0129) & 0.7328 (0.0896) & 0.9355 (0.0628) & 1.0000 (0.0000) \\
        \bottomrule
    \end{tabular}
\end{table*}

\begin{table*}
    \centering
    \caption{%
        Results on the held-out test set for TEX @ 125~Hz, varying $n$ and the authentication round.
        Values are presented as mean (standard deviation) across the 4~models trained with 4-fold cross-validation.%
    }
    \label{tab:tex125}
    \begin{tabular}{lcccccc}
        \toprule
        \multirow{2}[2]{*}{$n$} & \multirow{2}[2]{*}{Round} & \multirow{2}[2]{*}{EER} & \multicolumn{4}{c}{FRR @ FAR} \\\cmidrule{4-7}
        {} & {} & {} & $10^{-1}$ & $10^{-2}$ & $10^{-3}$ & $10^{-4}$ \\
        \midrule
        \multirow{9}{*}{1} & 1 & 0.3154 (0.0122) & 0.6525 (0.0407) & 0.9219 (0.0295) & 0.9799 (0.0117) & 0.9929 (0.0050) \\
        &      2 & 0.3395 (0.0164) & 0.6606 (0.0451) & 0.9485 (0.0388) & 0.9950 (0.0047) & 0.9994 (0.0008) \\
        &      3 & 0.3211 (0.0082) & 0.6277 (0.0301) & 0.9228 (0.0317) & 0.9870 (0.0145) & 0.9984 (0.0023) \\
        &      4 & 0.3339 (0.0248) & 0.6539 (0.0224) & 0.9408 (0.0345) & 0.9936 (0.0077) & 0.9987 (0.0020) \\
        &      5 & 0.3287 (0.0223) & 0.6462 (0.0723) & 0.9158 (0.0420) & 0.9834 (0.0154) & 0.9938 (0.0063) \\
        &      6 & 0.3489 (0.0145) & 0.6925 (0.0347) & 0.9265 (0.0228) & 0.9830 (0.0139) & 0.9940 (0.0081) \\
        &      7 & 0.3579 (0.0343) & 0.6300 (0.0533) & 0.8763 (0.0369) & 0.9662 (0.0329) & 0.9888 (0.0195) \\
        &      8 & 0.3215 (0.0227) & 0.6361 (0.0556) & 0.8966 (0.0390) & 0.9797 (0.0236) & 0.9892 (0.0159) \\
        &      9 & 0.3189 (0.0334) & 0.5440 (0.0250) & 0.8409 (0.0538) & 0.9980 (0.0035) & 1.0000 (0.0000) \\\cmidrule{2-7}
        \multirow{9}{*}{5} & 1 & 0.2038 (0.0122) & 0.3544 (0.0302) & 0.6818 (0.0207) & 0.8233 (0.0302) & 0.8879 (0.0461) \\
        &      2 & 0.2320 (0.0071) & 0.4594 (0.0346) & 0.8175 (0.0107) & 0.9222 (0.0214) & 0.9588 (0.0202) \\
        &      3 & 0.2328 (0.0158) & 0.4283 (0.0292) & 0.7578 (0.0338) & 0.8750 (0.0367) & 0.9212 (0.0392) \\
        &      4 & 0.2285 (0.0115) & 0.4404 (0.0332) & 0.8015 (0.0230) & 0.9125 (0.0346) & 0.9455 (0.0345) \\
        &      5 & 0.2598 (0.0084) & 0.5283 (0.0400) & 0.8159 (0.0325) & 0.9069 (0.0304) & 0.9384 (0.0304) \\
        &      6 & 0.2521 (0.0106) & 0.4895 (0.0205) & 0.8238 (0.0337) & 0.9247 (0.0353) & 0.9544 (0.0317) \\
        &      7 & 0.2499 (0.0240) & 0.4603 (0.0529) & 0.7631 (0.0471) & 0.8674 (0.0441) & 0.9021 (0.0463) \\
        &      8 & 0.2363 (0.0325) & 0.4536 (0.0492) & 0.7673 (0.0435) & 0.8831 (0.0309) & 0.9330 (0.0177) \\
        &      9 & 0.2492 (0.0105) & 0.4163 (0.0437) & 0.8338 (0.1113) & 0.9598 (0.0553) & 0.9802 (0.0310) \\\cmidrule{2-7}
        \multirow{9}{*}{10} & 1 & 0.1875 (0.0106) & 0.3123 (0.0281) & 0.6327 (0.0277) & 0.7990 (0.0276) & 0.8642 (0.0289) \\
       &      2 & 0.2207 (0.0130) & 0.4101 (0.0234) & 0.7665 (0.0184) & 0.9084 (0.0239) & 0.9578 (0.0257) \\
       &      3 & 0.2382 (0.0080) & 0.4492 (0.0221) & 0.7765 (0.0207) & 0.8913 (0.0335) & 0.9291 (0.0415) \\
       &      4 & 0.2059 (0.0069) & 0.3833 (0.0286) & 0.7414 (0.0128) & 0.8653 (0.0334) & 0.9050 (0.0359) \\
       &      5 & 0.2402 (0.0105) & 0.4715 (0.0141) & 0.7641 (0.0110) & 0.8703 (0.0133) & 0.9080 (0.0129) \\
       &      6 & 0.2499 (0.0165) & 0.4780 (0.0359) & 0.7994 (0.0335) & 0.9159 (0.0219) & 0.9504 (0.0201) \\
       &      7 & 0.2375 (0.0191) & 0.4337 (0.0473) & 0.7426 (0.0399) & 0.8594 (0.0329) & 0.9065 (0.0396) \\
       &      8 & 0.2420 (0.0369) & 0.4594 (0.0391) & 0.7764 (0.0198) & 0.9066 (0.0321) & 0.9695 (0.0271) \\
       &      9 & 0.2441 (0.0070) & 0.3937 (0.0167) & 0.6745 (0.0358) & 0.8276 (0.0261) & 0.9086 (0.0294) \\
        \bottomrule
    \end{tabular}
\end{table*}

\begin{table*}
    \centering
    \caption{%
        Results on the held-out test set for TEX @ 50~Hz, varying $n$ and the authentication round.
        Values are presented as mean (standard deviation) across the 4~models trained with 4-fold cross-validation.%
    }
    \label{tab:tex50}
    \begin{tabular}{lcccccc}
        \toprule
        \multirow{2}[2]{*}{$n$} & \multirow{2}[2]{*}{Round} & \multirow{2}[2]{*}{EER} & \multicolumn{4}{c}{FRR @ FAR} \\\cmidrule{4-7}
        {} & {} & {} & $10^{-1}$ & $10^{-2}$ & $10^{-3}$ & $10^{-4}$ \\
        \midrule
        \multirow{9}{*}{1} & 1 & 0.3864 (0.0171) & 0.7493 (0.0264) & 0.9443 (0.0381) & 0.9701 (0.0350) & 0.9724 (0.0325) \\
        &      2 & 0.3752 (0.0402) & 0.7336 (0.0541) & 0.9315 (0.0449) & 0.9509 (0.0386) & 0.9533 (0.0375) \\
        &      3 & 0.3974 (0.0126) & 0.7902 (0.0255) & 0.9552 (0.0449) & 0.9659 (0.0344) & 0.9674 (0.0330) \\
        &      4 & 0.3830 (0.0138) & 0.7611 (0.0383) & 0.9138 (0.0354) & 0.9366 (0.0318) & 0.9404 (0.0304) \\
        &      5 & 0.3695 (0.0261) & 0.7485 (0.1031) & 1.0000 (0.0000) & 1.0000 (0.0000) & 1.0000 (0.0000) \\
        &      6 & 0.3764 (0.0145) & 0.7345 (0.0188) & 0.8811 (0.0334) & 0.9042 (0.0434) & 0.9085 (0.0472) \\
        &      7 & 0.3721 (0.0594) & 0.7413 (0.0643) & 0.9509 (0.0394) & 0.9746 (0.0280) & 0.9772 (0.0249) \\
        &      8 & 0.3111 (0.0206) & 0.5650 (0.0505) & 0.8874 (0.1138) & 0.9416 (0.0845) & 0.9514 (0.0842) \\
        &      9 & 0.3351 (0.0441) & 0.6107 (0.1639) & 0.8460 (0.1657) & 1.0000 (0.0000) & 1.0000 (0.0000) \\\cmidrule{2-7}
        \multirow{9}{*}{5} & 1 & 0.2695 (0.0130) & 0.4923 (0.0220) & 0.6525 (0.0313) & 0.6748 (0.0342) & 0.6787 (0.0354) \\
        &      2 & 0.2688 (0.0072) & 0.5291 (0.0342) & 0.7228 (0.0505) & 0.7519 (0.0556) & 0.7573 (0.0571) \\
        &      3 & 0.2677 (0.0096) & 0.5479 (0.0219) & 0.7490 (0.0125) & 0.7768 (0.0100) & 0.7815 (0.0103) \\
        &      4 & 0.2754 (0.0100) & 0.5597 (0.0273) & 0.7497 (0.0330) & 0.7764 (0.0387) & 0.7800 (0.0398) \\
        &      5 & 0.2889 (0.0158) & 0.6037 (0.0480) & 0.7814 (0.0418) & 0.8039 (0.0401) & 0.8079 (0.0396) \\
        &      6 & 0.2769 (0.0097) & 0.5556 (0.0295) & 0.7497 (0.0284) & 0.7805 (0.0320) & 0.7865 (0.0342) \\
        &      7 & 0.2993 (0.0151) & 0.5426 (0.0496) & 0.7098 (0.0418) & 0.7408 (0.0443) & 0.7445 (0.0448) \\
        &      8 & 0.2860 (0.0068) & 0.6099 (0.0336) & 0.8199 (0.0335) & 0.8536 (0.0333) & 0.8608 (0.0334) \\
        &      9 & 0.3233 (0.0400) & 0.6066 (0.1032) & 0.7753 (0.1035) & 0.8072 (0.1138) & 0.8149 (0.1172) \\\cmidrule{2-7}
        \multirow{9}{*}{10} & 1 & 0.2371 (0.0436) & 0.4152 (0.1301) & 0.5411 (0.1669) & 0.5558 (0.1720) & 0.5577 (0.1730) \\
       &      2 & 0.2423 (0.0316) & 0.4523 (0.1183) & 0.6164 (0.1602) & 0.6369 (0.1674) & 0.6387 (0.1681) \\
       &      3 & 0.2407 (0.0373) & 0.4441 (0.1229) & 0.7699 (0.1339) & 0.7857 (0.1250) & 0.7876 (0.1239) \\
       &      4 & 0.2410 (0.0336) & 0.4605 (0.1244) & 0.6715 (0.0740) & 0.8032 (0.1139) & 0.8052 (0.1128) \\
       &      5 & 0.2580 (0.0112) & 0.5115 (0.0525) & 0.6638 (0.0710) & 0.6790 (0.0747) & 0.6806 (0.0754) \\
       &      6 & 0.2466 (0.0394) & 0.4742 (0.1329) & 0.7981 (0.1169) & 0.8150 (0.1075) & 0.8173 (0.1064) \\
       &      7 & 0.2673 (0.0330) & 0.4740 (0.1121) & 0.6143 (0.1267) & 0.6334 (0.1280) & 0.6354 (0.1287) \\
       &      8 & 0.2507 (0.0099) & 0.5058 (0.0858) & 0.7139 (0.1221) & 0.7486 (0.1250) & 0.7552 (0.1267) \\
       &      9 & 0.2664 (0.0663) & 0.4482 (0.1679) & 0.5627 (0.1842) & 0.5966 (0.1585) & 0.7664 (0.1380) \\
        \bottomrule
    \end{tabular}
\end{table*}

\begin{table*}
    \centering
    \caption{%
        Results on the held-out test set for TEX @ 31.25~Hz, varying $n$ and the authentication round.
        Values are presented as mean (standard deviation) across the 4~models trained with 4-fold cross-validation.%
    }
    \label{tab:tex31}
    \begin{tabular}{lcccccc}
        \toprule
        \multirow{2}[2]{*}{$n$} & \multirow{2}[2]{*}{Round} & \multirow{2}[2]{*}{EER} & \multicolumn{4}{c}{FRR @ FAR} \\\cmidrule{4-7}
        {} & {} & {} & $10^{-1}$ & $10^{-2}$ & $10^{-3}$ & $10^{-4}$ \\
        \midrule
        \multirow{9}{*}{1} & 1 & 0.4092 (0.0103) & 0.7694 (0.0389) & 0.9675 (0.0329) & 0.9904 (0.0117) & 0.9944 (0.0096) \\
        &      2 & 0.4481 (0.0198) & 0.8507 (0.0441) & 0.9740 (0.0111) & 0.9937 (0.0042) & 0.9986 (0.0012) \\
        &      3 & 0.4456 (0.0092) & 0.8727 (0.0739) & 0.9696 (0.0284) & 0.9823 (0.0203) & 0.9846 (0.0186) \\
        &      4 & 0.4493 (0.0419) & 0.8455 (0.0417) & 0.9918 (0.0060) & 1.0000 (0.0000) & 1.0000 (0.0000) \\
        &      5 & 0.4516 (0.0018) & 0.8459 (0.0359) & 0.9688 (0.0202) & 0.9898 (0.0089) & 0.9941 (0.0057) \\
        &      6 & 0.4429 (0.0073) & 0.8631 (0.0313) & 0.9772 (0.0151) & 0.9933 (0.0039) & 0.9970 (0.0027) \\
        &      7 & 0.4370 (0.0376) & 0.8467 (0.0419) & 0.9886 (0.0124) & 0.9980 (0.0020) & 0.9994 (0.0009) \\
        &      8 & 0.4446 (0.0339) & 0.8506 (0.0542) & 0.9807 (0.0307) & 0.9832 (0.0286) & 0.9836 (0.0284) \\
        &      9 & 0.3834 (0.0174) & 0.7227 (0.0830) & 0.9768 (0.0244) & 0.9880 (0.0175) & 0.9900 (0.0174) \\\cmidrule{2-7}
        \multirow{9}{*}{5} & 1 & 0.2919 (0.0124) & 0.5907 (0.0351) & 0.8395 (0.0240) & 0.9095 (0.0355) & 0.9284 (0.0343) \\
        &      2 & 0.3321 (0.0114) & 0.6685 (0.0215) & 0.8896 (0.0298) & 0.9391 (0.0386) & 0.9511 (0.0367) \\
        &      3 & 0.3350 (0.0363) & 0.6593 (0.0552) & 0.8677 (0.0369) & 0.9207 (0.0443) & 0.9363 (0.0499) \\
        &      4 & 0.3553 (0.0158) & 0.6812 (0.0162) & 0.8453 (0.0268) & 0.8869 (0.0358) & 0.8995 (0.0385) \\
        &      5 & 0.3709 (0.0218) & 0.6939 (0.0244) & 0.8599 (0.0178) & 0.9048 (0.0311) & 0.9190 (0.0365) \\
        &      6 & 0.3171 (0.0350) & 0.6148 (0.0415) & 0.8234 (0.0568) & 0.8775 (0.0748) & 0.8905 (0.0800) \\
        &      7 & 0.3176 (0.0270) & 0.6134 (0.0284) & 0.8251 (0.0563) & 0.8829 (0.0731) & 0.9007 (0.0795) \\
        &      8 & 0.3555 (0.0315) & 0.6597 (0.0468) & 0.8193 (0.0452) & 0.8587 (0.0651) & 0.8687 (0.0715) \\
        &      9 & 0.3281 (0.0469) & 0.6559 (0.1135) & 0.9035 (0.0926) & 0.9460 (0.0572) & 0.9562 (0.0448) \\\cmidrule{2-7}
        \multirow{9}{*}{10} & 1 & 0.2666 (0.0246) & 0.5147 (0.0628) & 0.7356 (0.0476) & 0.7881 (0.0645) & 0.8079 (0.0779) \\
       &      2 & 0.3269 (0.0229) & 0.6328 (0.0161) & 0.8127 (0.0365) & 0.8566 (0.0573) & 0.8672 (0.0634) \\
       &      3 & 0.3175 (0.0228) & 0.6192 (0.0425) & 0.7960 (0.0273) & 0.8370 (0.0445) & 0.8506 (0.0535) \\
       &      4 & 0.3376 (0.0386) & 0.6358 (0.0601) & 0.7887 (0.0508) & 0.8230 (0.0602) & 0.8322 (0.0647) \\
       &      5 & 0.3480 (0.0185) & 0.6446 (0.0230) & 0.7766 (0.0287) & 0.8062 (0.0480) & 0.8151 (0.0567) \\
       &      6 & 0.3047 (0.0281) & 0.5902 (0.0591) & 0.7796 (0.0368) & 0.8282 (0.0537) & 0.8404 (0.0600) \\
       &      7 & 0.3218 (0.0233) & 0.6363 (0.0436) & 0.7977 (0.0437) & 0.8400 (0.0620) & 0.8469 (0.0659) \\
       &      8 & 0.3367 (0.0391) & 0.6416 (0.0577) & 0.8069 (0.0192) & 0.8522 (0.0401) & 0.8692 (0.0538) \\
       &      9 & 0.3205 (0.0443) & 0.6299 (0.1492) & 0.8001 (0.1110) & 0.8538 (0.0781) & 0.8762 (0.0609) \\
        \bottomrule
    \end{tabular}
\end{table*}

\begin{table*}
    \centering
    \caption{%
        Results on the held-out test set for the DEL baseline on TEX @ 1000~Hz, varying $n$ and the authentication round.
        Values are presented as mean (standard deviation) across the 4~models trained with 4-fold cross-validation.%
    }
    \label{tab:del}
    
    \rev{
    \begin{tabular}{lcccccc}
        \toprule
        \multirow{2}[2]{*}{$n$} & \multirow{2}[2]{*}{Round} & \multirow{2}[2]{*}{EER} & \multicolumn{4}{c}{FRR @ FAR} \\\cmidrule{4-7}
        {} & {} & {} & $10^{-1}$ & $10^{-2}$ & $10^{-3}$ & $10^{-4}$ \\
        \midrule
    
       \multirow{9}{*}{1} &      1 & 0.4641 (0.0101) & 0.8793 (0.0540) & 0.9814 (0.0239) & 0.9890 (0.0190) & 0.9896 (0.0181) \\
       {} &      2 & 0.4783 (0.0057) & 0.8652 (0.0268) & 0.9638 (0.0144) & 0.9797 (0.0127) & 0.9828 (0.0130) \\
       {} &      3 & 0.4857 (0.0119) & 0.8849 (0.0245) & 0.9839 (0.0095) & 0.9958 (0.0043) & 0.9973 (0.0030) \\
       {} &      4 & 0.4882 (0.0341) & 0.9068 (0.0393) & 0.9951 (0.0084) & 0.9967 (0.0058) & 0.9968 (0.0056) \\
       {} &      5 & 0.4801 (0.0212) & 0.8964 (0.0061) & 0.9910 (0.0157) & 0.9939 (0.0106) & 0.9944 (0.0097) \\
       {} &      6 & 0.4793 (0.0245) & 0.8743 (0.0281) & 0.9772 (0.0100) & 0.9905 (0.0103) & 0.9923 (0.0105) \\
       {} &      7 & 0.5044 (0.0256) & 0.8839 (0.0133) & 0.9779 (0.0147) & 0.9894 (0.0107) & 0.9913 (0.0102) \\
       {} &      8 & 0.5182 (0.0162) & 0.9240 (0.0179) & 0.9846 (0.0146) & 0.9895 (0.0109) & 0.9907 (0.0101) \\
       {} &      9 & 0.4804 (0.0504) & 0.8303 (0.0693) & 0.9361 (0.0357) & 0.9562 (0.0282) & 0.9599 (0.0266) \\\cmidrule{2-7}
       \multirow{9}{*}{5} &      1 & 0.4332 (0.0191) & 0.7945 (0.0091) & 0.9723 (0.0133) & 0.9965 (0.0039) & 0.9992 (0.0010) \\
       {} &      2 & 0.4684 (0.0186) & 0.8608 (0.0367) & 0.9863 (0.0166) & 0.9984 (0.0026) & 1.0000 (0.0000) \\
       {} &      3 & 0.4578 (0.0204) & 0.8388 (0.0405) & 0.9834 (0.0178) & 0.9957 (0.0066) & 0.9977 (0.0037) \\
       {} &      4 & 0.4821 (0.0338) & 0.8477 (0.0475) & 0.9762 (0.0331) & 0.9872 (0.0201) & 0.9912 (0.0145) \\
       {} &      5 & 0.4531 (0.0177) & 0.8482 (0.0201) & 0.9757 (0.0148) & 0.9930 (0.0056) & 0.9978 (0.0021) \\
       {} &      6 & 0.4817 (0.0103) & 0.8756 (0.0142) & 0.9805 (0.0079) & 0.9960 (0.0021) & 0.9988 (0.0009) \\
       {} &      7 & 0.5148 (0.0264) & 0.9110 (0.0536) & 0.9859 (0.0161) & 0.9947 (0.0075) & 0.9968 (0.0048) \\
       {} &      8 & 0.4679 (0.0504) & 0.8416 (0.0761) & 0.9838 (0.0176) & 0.9935 (0.0080) & 0.9969 (0.0037) \\
       {} &      9 & 0.5350 (0.0318) & 0.9034 (0.0857) & 0.9893 (0.0183) & 1.0000 (0.0000) & 1.0000 (0.0000) \\\cmidrule{2-7}
      \multirow{9}{*}{10} &      1 & 0.4295 (0.0213) & 0.8001 (0.0240) & 0.9706 (0.0165) & 0.9955 (0.0033) & 0.9992 (0.0005) \\
      {} &      2 & 0.4590 (0.0191) & 0.8375 (0.0395) & 0.9706 (0.0194) & 0.9943 (0.0066) & 0.9991 (0.0009) \\
      {} &      3 & 0.4490 (0.0141) & 0.8516 (0.0428) & 0.9866 (0.0124) & 0.9988 (0.0016) & 0.9998 (0.0003) \\
      {} &      4 & 0.4544 (0.0205) & 0.8253 (0.0474) & 0.9716 (0.0189) & 0.9949 (0.0072) & 0.9990 (0.0017) \\
      {} &      5 & 0.4730 (0.0259) & 0.8213 (0.0132) & 0.9459 (0.0150) & 0.9755 (0.0153) & 0.9849 (0.0120) \\
      {} &      6 & 0.4748 (0.0197) & 0.8465 (0.0301) & 0.9867 (0.0172) & 0.9973 (0.0034) & 0.9991 (0.0010) \\
      {} &      7 & 0.5010 (0.0233) & 0.9314 (0.0424) & 0.9936 (0.0064) & 0.9984 (0.0017) & 0.9995 (0.0006) \\
      {} &      8 & 0.4967 (0.0378) & 0.8710 (0.0672) & 0.9956 (0.0044) & 0.9996 (0.0007) & 0.9999 (0.0001) \\
      {} &      9 & 0.5164 (0.0239) & 0.9239 (0.0520) & 0.9955 (0.0073) & 0.9997 (0.0006) & 1.0000 (0.0000) \\
        \bottomrule
    \end{tabular}
    }
\end{table*}

\begin{table*}
\centering
\rev{
\caption{%
    Results on the held-out test set for the DEL (slow subnet only) baseline on TEX @ 1000~Hz, varying $n$ and the authentication round.
    Values are presented as mean (standard deviation) across the 4~models trained with 4-fold cross-validation.%
}
\label{tab:del-slow}
}

\rev{
\begin{tabular}{lcccccc}
\toprule
\multirow{2}[2]{*}{$n$} & \multirow{2}[2]{*}{Round} & \multirow{2}[2]{*}{EER} & \multicolumn{4}{c}{FRR @ FAR} \\\cmidrule{4-7}
        {} & {} & {} & $10^{-1}$ & $10^{-2}$ & $10^{-3}$ & $10^{-4}$ \\
\midrule
       \multirow{9}{*}{1} &      1 & 0.2332 (0.0281) & 0.4115 (0.0647) & 0.8396 (0.0965) & 0.9708 (0.0294) & 0.9948 (0.0065) \\
        &      2 & 0.3253 (0.0278) & 0.6551 (0.0812) & 0.9255 (0.0408) & 0.9898 (0.0095) & 0.9977 (0.0035) \\
        &      3 & 0.3311 (0.0471) & 0.6345 (0.0584) & 0.9435 (0.0522) & 0.9886 (0.0141) & 0.9971 (0.0045) \\
        &      4 & 0.3389 (0.0310) & 0.6288 (0.0348) & 0.9285 (0.0487) & 0.9861 (0.0170) & 0.9986 (0.0015) \\
        &      5 & 0.3037 (0.0442) & 0.5702 (0.0772) & 0.8724 (0.0406) & 0.9750 (0.0202) & 0.9965 (0.0050) \\
        &      6 & 0.3293 (0.0164) & 0.6287 (0.0726) & 0.9331 (0.0642) & 0.9902 (0.0098) & 0.9997 (0.0005) \\
        &      7 & 0.3395 (0.0241) & 0.6411 (0.0515) & 0.9243 (0.0532) & 0.9892 (0.0135) & 0.9988 (0.0019) \\
        &      8 & 0.3170 (0.0283) & 0.5862 (0.0760) & 0.9119 (0.0652) & 0.9973 (0.0047) & 0.9993 (0.0013) \\
        &      9 & 0.3624 (0.0119) & 0.6119 (0.0299) & 0.8334 (0.0726) & 0.9218 (0.1030) & 0.9371 (0.1079) \\ \cmidrule{2-7}
       \multirow{9}{*}{5} &      1 & 0.1660 (0.0397) & 0.2495 (0.0937) & 0.6315 (0.1347) & 0.8590 (0.0995) & 0.9334 (0.0515) \\
        &      2 & 0.2203 (0.0492) & 0.3883 (0.0987) & 0.8148 (0.0432) & 0.9682 (0.0197) & 0.9979 (0.0021) \\
        &      3 & 0.2503 (0.0447) & 0.4331 (0.0900) & 0.7924 (0.0961) & 0.9544 (0.0434) & 0.9890 (0.0121) \\
        &      4 & 0.2449 (0.0550) & 0.4299 (0.1031) & 0.8121 (0.1135) & 0.9409 (0.0386) & 0.9786 (0.0175) \\
        &      5 & 0.2216 (0.0456) & 0.4017 (0.0912) & 0.7537 (0.0661) & 0.8948 (0.0299) & 0.9588 (0.0128) \\
        &      6 & 0.2360 (0.0245) & 0.4258 (0.0768) & 0.7763 (0.0907) & 0.9326 (0.0493) & 0.9866 (0.0178) \\
        &      7 & 0.2683 (0.0365) & 0.4962 (0.0832) & 0.8135 (0.0587) & 0.9159 (0.0347) & 0.9510 (0.0281) \\
        &      8 & 0.2449 (0.0462) & 0.4449 (0.0965) & 0.8142 (0.0483) & 0.9578 (0.0284) & 0.9953 (0.0072) \\
        &      9 & 0.2745 (0.0382) & 0.5231 (0.0619) & 0.8535 (0.0688) & 0.9446 (0.0373) & 0.9744 (0.0236) \\ \cmidrule{2-7}
      \multirow{9}{*}{10} &      1 & 0.1559 (0.0394) & 0.2226 (0.0854) & 0.5822 (0.1083) & 0.8558 (0.1072) & 0.9550 (0.0483) \\
       &      2 & 0.2151 (0.0457) & 0.3669 (0.0962) & 0.7724 (0.0386) & 0.9583 (0.0245) & 0.9941 (0.0063) \\
       &      3 & 0.2421 (0.0477) & 0.4139 (0.1006) & 0.7571 (0.0959) & 0.9362 (0.0650) & 0.9909 (0.0139) \\
       &      4 & 0.2261 (0.0549) & 0.3873 (0.1068) & 0.7729 (0.0664) & 0.9554 (0.0405) & 0.9906 (0.0108) \\
       &      5 & 0.2070 (0.0415) & 0.3634 (0.0823) & 0.7431 (0.0821) & 0.8780 (0.0593) & 0.9395 (0.0282) \\
       &      6 & 0.2314 (0.0336) & 0.4053 (0.0807) & 0.7619 (0.0662) & 0.9241 (0.0438) & 0.9687 (0.0230) \\
       &      7 & 0.2569 (0.0354) & 0.4523 (0.0716) & 0.7660 (0.0482) & 0.9119 (0.0367) & 0.9666 (0.0239) \\
       &      8 & 0.2378 (0.0384) & 0.4444 (0.0840) & 0.8274 (0.0464) & 0.9675 (0.0192) & 0.9912 (0.0096) \\
       &      9 & 0.2512 (0.0277) & 0.4875 (0.0485) & 0.8387 (0.1023) & 0.9257 (0.0512) & 0.9600 (0.0278) \\
\bottomrule
\end{tabular}
}
\end{table*}

\begin{table*}
\centering
\rev{
\caption{%
    Results on the held-out test set for the DEL (fast subnet only) baseline on TEX @ 1000~Hz, varying $n$ and the authentication round.
    Values are presented as mean (standard deviation) across the 4~models trained with 4-fold cross-validation.%
}
\label{tab:del-fast}
}

\rev{
\begin{tabular}{lcccccc}
\toprule
\multirow{2}[2]{*}{$n$} & \multirow{2}[2]{*}{Round} & \multirow{2}[2]{*}{EER} & \multicolumn{4}{c}{FRR @ FAR} \\\cmidrule{4-7}
        {} & {} & {} & $10^{-1}$ & $10^{-2}$ & $10^{-3}$ & $10^{-4}$ \\
\midrule
       \multirow{9}{*}{1} &      1 & 0.3221 (0.0280) & 0.5954 (0.0483) & 0.9365 (0.0642) & 0.9948 (0.0058) & 1.0000 (0.0000) \\
        &      2 & 0.3575 (0.0119) & 0.7035 (0.0331) & 0.9530 (0.0316) & 0.9914 (0.0149) & 0.9929 (0.0124) \\
        &      3 & 0.3823 (0.0191) & 0.7518 (0.0725) & 0.9660 (0.0322) & 0.9951 (0.0085) & 0.9975 (0.0043) \\
        &      4 & 0.4091 (0.0149) & 0.7863 (0.0227) & 0.9667 (0.0204) & 0.9935 (0.0065) & 0.9980 (0.0029) \\
        &      5 & 0.3458 (0.0264) & 0.6753 (0.0321) & 0.9596 (0.0405) & 0.9899 (0.0109) & 0.9957 (0.0056) \\
        &      6 & 0.3975 (0.0165) & 0.7941 (0.0152) & 0.9839 (0.0092) & 0.9996 (0.0006) & 1.0000 (0.0000) \\
        &      7 & 0.3596 (0.0051) & 0.6975 (0.0344) & 0.9751 (0.0251) & 0.9959 (0.0071) & 0.9973 (0.0046) \\
        &      8 & 0.3599 (0.0278) & 0.6587 (0.0498) & 0.9561 (0.0316) & 0.9996 (0.0007) & 1.0000 (0.0000) \\
        &      9 & 0.3439 (0.0437) & 0.6145 (0.0823) & 0.8998 (0.0801) & 0.9657 (0.0376) & 0.9828 (0.0241) \\ \cmidrule{2-7}
       \multirow{9}{*}{5} &      1 & 0.2297 (0.0417) & 0.4059 (0.0817) & 0.7491 (0.0323) & 0.8799 (0.0371) & 0.9170 (0.0466) \\
        &      2 & 0.2763 (0.0327) & 0.5209 (0.0391) & 0.8540 (0.0096) & 0.9738 (0.0267) & 0.9851 (0.0153) \\
        &      3 & 0.3124 (0.0325) & 0.5915 (0.0641) & 0.8966 (0.0712) & 0.9659 (0.0297) & 0.9832 (0.0174) \\
        &      4 & 0.3071 (0.0323) & 0.6142 (0.0674) & 0.8982 (0.0519) & 0.9620 (0.0227) & 0.9779 (0.0137) \\
        &      5 & 0.2856 (0.0378) & 0.5704 (0.0938) & 0.8944 (0.0744) & 0.9573 (0.0354) & 0.9746 (0.0228) \\
        &      6 & 0.2754 (0.0330) & 0.5628 (0.0640) & 0.8758 (0.0172) & 0.9497 (0.0143) & 0.9659 (0.0223) \\
        &      7 & 0.2969 (0.0323) & 0.5818 (0.0670) & 0.8560 (0.0710) & 0.9341 (0.0444) & 0.9682 (0.0496) \\
        &      8 & 0.2886 (0.0194) & 0.5423 (0.0376) & 0.8947 (0.0632) & 0.9796 (0.0136) & 0.9995 (0.0008) \\
        &      9 & 0.3268 (0.0382) & 0.6206 (0.0759) & 0.9406 (0.0697) & 0.9757 (0.0420) & 0.9790 (0.0364) \\ \cmidrule{2-7}
      \multirow{9}{*}{10} &      1 & 0.2111 (0.0377) & 0.3631 (0.0737) & 0.6995 (0.0371) & 0.8208 (0.0766) & 0.8691 (0.0973) \\
       &      2 & 0.2680 (0.0338) & 0.5113 (0.0440) & 0.7994 (0.0302) & 0.8729 (0.0543) & 0.8927 (0.0622) \\
       &      3 & 0.3020 (0.0326) & 0.5582 (0.0489) & 0.8793 (0.0845) & 0.9504 (0.0535) & 0.9620 (0.0421) \\
       &      4 & 0.2743 (0.0359) & 0.5366 (0.0733) & 0.8734 (0.0873) & 0.9458 (0.0503) & 0.9690 (0.0437) \\
       &      5 & 0.2714 (0.0345) & 0.5234 (0.0647) & 0.8862 (0.0930) & 0.9460 (0.0577) & 0.9602 (0.0454) \\
       &      6 & 0.2673 (0.0326) & 0.5296 (0.0612) & 0.8208 (0.0189) & 0.8967 (0.0317) & 0.9146 (0.0403) \\
       &      7 & 0.2778 (0.0347) & 0.5389 (0.0406) & 0.8451 (0.0660) & 0.9263 (0.0708) & 0.9380 (0.0677) \\
       &      8 & 0.2742 (0.0256) & 0.5296 (0.0413) & 0.9459 (0.0493) & 0.9879 (0.0206) & 0.9918 (0.0142) \\
       &      9 & 0.3208 (0.0161) & 0.5986 (0.0782) & 0.8478 (0.0555) & 0.9610 (0.0474) & 0.9647 (0.0456) \\
\bottomrule
\end{tabular}
}
\end{table*}

\begin{table*}
    \centering
    \caption{%
        Results on the held-out test set for the STAR baseline on TEX @ 1000~Hz, varying the authentication round.
        Note that STAR uses the full recording duration.
        Values are presented as mean (standard deviation) across the 4~models trained with 4-fold cross-validation.%
    }
    \label{tab:star}
    \begin{tabular}{cccccc}
        \toprule
        \multirow{2}[2]{*}{Round} & \multirow{2}[2]{*}{EER} & \multicolumn{4}{c}{FRR @ FAR} \\\cmidrule{3-6}
        {} & {} & $10^{-1}$ & $10^{-2}$ & $10^{-3}$ & $10^{-4}$ \\
        \midrule
        1 & 0.1563 (0.0487) & 0.2249 (0.1101) & 0.6240 (0.2479) & 0.8040 (0.1962) & 0.9107 (0.0917) \\
        2 & 0.2236 (0.0432) & 0.4133 (0.1297) & 0.8105 (0.1901) & 0.8995 (0.1010) & 0.9338 (0.0662) \\
        3 & 0.2415 (0.0537) & 0.4274 (0.1339) & 0.8077 (0.1964) & 0.9012 (0.1020) & 0.9377 (0.0636) \\
        4 & 0.2243 (0.0387) & 0.4406 (0.1519) & 0.8164 (0.1852) & 0.9106 (0.0899) & 0.9503 (0.0514) \\
        5 & 0.2193 (0.0481) & 0.4200 (0.1460) & 0.8034 (0.1614) & 0.9116 (0.0903) & 0.9624 (0.0388) \\
        6 & 0.2461 (0.0620) & 0.4161 (0.1188) & 0.8284 (0.1727) & 0.9237 (0.0765) & 0.9727 (0.0281) \\
        7 & 0.4831 (0.0242) & 0.8872 (0.0080) & 0.9665 (0.0384) & 0.9756 (0.0400) & 0.9765 (0.0405) \\
        8 & 0.4397 (0.0446) & 0.8165 (0.0359) & 0.9818 (0.0226) & 0.9957 (0.0062) & 0.9989 (0.0017) \\
        9 & 0.5051 (0.0159) & 0.8500 (0.1070) & 0.9309 (0.1139) & 0.9335 (0.1151) & 0.9335 (0.1151) \\
        \bottomrule
    \end{tabular}
\end{table*}

\end{document}